%% file: main.tex
\documentclass[a4paper, twoside, openright]{report}

\usepackage[a4paper,top=3cm,bottom=3cm,left=3.5cm,right=3.5cm,heightrounded]{geometry}
\usepackage[fontsize=13pt]{scrextend}
\usepackage[english]{babel}
\usepackage[fixlanguage]{babelbib} 
\usepackage[utf8]{inputenc} 
\usepackage[T1]{fontenc} 

\usepackage{lipsum} 
\usepackage{fancyhdr} 
\usepackage{titlesec} 
\usepackage[normalem]{ulem} 
\usepackage[dvipsnames]{xcolor} 
\usepackage{graphicx} 
\usepackage{rotating} 
\usepackage{listings} 
\usepackage{hyperref} 
\usepackage{xurl} 
\usepackage{svg}

\usepackage{caption}
\usepackage{subcaption}

\usepackage{amssymb}
\usepackage{amsmath}
\usepackage{amsthm}

\hypersetup{hidelinks} 

\input{configs/headers}
\input{configs/code_style}
\input{configs/footnotes}

\newenvironment{acknowledgements} {\begin{abstract}} {\end{abstract}}

\newcommand{\CSharp}{\texttt{C\#}}
\newcommand{\FSharp}{\texttt{F\#}}

\begin{document}

\include{chapters/title_page}
\include{chapters/acknowledgements}
\include{chapters/abstract}
\cleardoublepage

\tableofcontents

\titleformat{\chapter}{\normalfont\fontsize{30}{30}\bfseries}{\thechapter.}{0.5em}{}
\include{chapters/chapter_1/chapter_1}
\include{chapters/chapter_2/chapter_2}
\include{chapters/chapter_3/chapter_3}
\include{chapters/chapter_4/chapter_4}
\include{chapters/chapter_5/chapter_5}


\bibliographystyle{ieeetr}
\bibliography{chapters/bibliography}

\end{document}

%% file: configs/headers.tex
\fancypagestyle{plain}{ 
  \fancyhf{}
  
}

%% file: configs/code_style.tex
\lstdefinestyle{codeStyle}{
    commentstyle=\color{teal},
    keywordstyle=\color{Magenta},
    numberstyle=\tiny\color{gray},
    stringstyle=\color{violet},
    basicstyle=\ttfamily\footnotesize,
    breakatwhitespace=false,     
    breaklines=true,                 
    captionpos=b,                    
    keepspaces=true,                 
    numbers=left,                    
    numbersep=5pt,                  
    showspaces=false,                
    showstringspaces=false,
    showtabs=false,
    tabsize=2
} 

\lstdefinestyle{longBlock}{
    commentstyle=\color{teal},
    keywordstyle=\color{Magenta},
    numberstyle=\tiny\color{gray},
    stringstyle=\color{violet},
    basicstyle=\ttfamily\scriptsize,
    breakatwhitespace=false,         
    breaklines=true,                 
    captionpos=b,                    
    keepspaces=true,                 
    numbers=left,                    
    numbersep=5pt,                  
    showspaces=false,                
    showstringspaces=false,
    showtabs=false,                  
    tabsize=2
} 

%% file: configs/footnotes.tex
\deffootnote{\dimexpr1.5em+6mm}{1em}{\thefootnotemark.\hspace{3mm}}

%% file: chapters/title_page.tex
\begin{titlepage}
\begin{figure}[!htb]
    \centering
    \includegraphics[keepaspectratio=true,scale=0.5]{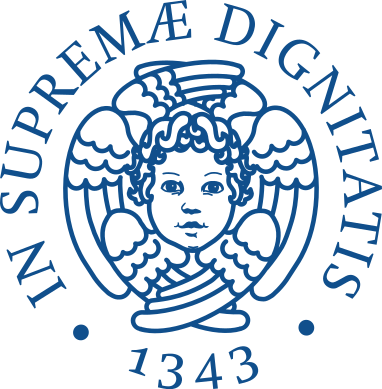}
\end{figure}

\begin{center}
    \LARGE{UNIVERSITÀ DI PISA}
    \vspace{3mm}
    \\ \large{DEPARTMENT OF COMPUTER SCIENCE}
    \vspace{3mm}
    \\ \LARGE{BSc in Computer Science}
\end{center}

\vspace{8mm}

\begin{center}
    {\LARGE{\textbf{Adaptive and Gamified Learning Paths\\ \vspace{3mm} with Polyglot and .NET Interactive }}}
    
    
\end{center}

\vspace{0.5mm}

\begin{center}
    \textit{by} \\ \vspace{3mm} \textbf{\large{Tommaso Martorella}}
\end{center}

\vspace{8mm}

\begin{center}
    \textit{supervisors} \\ \vspace{3mm}
    \textbf{\large{Dr. Antonio Bucchiarone}} \\ \vspace{2mm} \textbf{\large{Prof. Rossano Venturini}}
\end{center}

\vspace{30mm}
\centering{\large{ACADEMIC YEAR 2021/2022}}

\end{titlepage}

%% file: chapters/acknowledgements.tex
\begin{acknowledgements}

\vspace{1cm}
\noindent\hspace{0.0675\textwidth}
\begin{minipage}{0.865\textwidth}
\textit{
    First and foremost, I'm extremely grateful to my supervisor Antonio Bucchiarone, whose invaluable experience, constructive advice, and profound belief in my abilities made this work possible. I'd also like to extend my gratitude to the entire Fondazione Bruno Kessler for this incredible internship opportunity. Special thanks to Diego Colombo for introducing me to Antonio and for his irreplaceable insight, guidance, and support throughout my entire journey. To conclude, I cannot forget to thank my family and friends for all the unconditional support in this very intense academic year.
}
\end{minipage}
\end{acknowledgements}

%% file: chapters/abstract.tex
\begin{abstract}

\vspace{1cm}
\noindent\hspace{0.0675\textwidth}
\begin{minipage}{0.865\textwidth}
The digital age is changing the role of educators and pushing for a paradigm shift in the education system as a whole. Growing demand for general and specialized education inside and outside classrooms is at the heart of this rising trend. In modern, heterogeneous learning environments, the one-size-fits-all approach is proven to be fundamentally flawed. Individualization through adaptivity is, therefore, crucial to nurture individual potential and address accessibility needs and neurodiversity. By formalizing a learning framework that takes into account all these different aspects, we aim to define and implement an open, content-agnostic, and extensible platform to design and consume adaptive and gamified learning experiences.
\end{minipage}

\end{abstract}

%% file: chapters/chapter_1/chapter_1.tex
\chapter{Introduction}
\paragraph{}
\textbf{Adaptive learning} is the delivery of personalized learning experiences that address an individual's unique needs instead of a one-size-fits-all approach. It may be achieved through just-in-time feedback, personalized learning paths, ad-hoc resources, or another wide array of techniques. But why is it important?

\section{The education revolution}
\paragraph{}
Education is universally recognized as one of the factors with the highest impact on society and the individual. The United Nations included education in their 2030 Agenda for Sustainable Development\footnote{\url{https://www.un.org/sustainabledevelopment/education/}}. UNESCO launched the Global Education Coalition\footnote{\url{https://en.unesco.org/covid19/educationresponse/globalcoalition}} in response to the COVID-19 pandemic. The European Union created the Digital Education Action Plan (2021-2027)\footnote{\url{https://education.ec.europa.eu/focus-topics/digital-education/digital-education-action-plan}} to foster and support the adaptation of educational systems in the digital age. This collective global effort is motivated by a continuously increasing technology availability and a rising global enrolment rate. Furthermore, according to UNESCO, higher education is the fastest-growing sector\footnote{\input{chapters/chapter_1/footnote_1}}, with its global enrolment rate doubled in the last twenty years \cite{UNESCO_FLP}.

\paragraph{}
Rising trends such as \textbf{flexible learning pathways} and \textbf{micro-credentials}\footnote{ \url{https://education.ec.europa.eu/education-levels/higher-education/micro-credentials}} tend towards more versatile forms of content delivery and credential recognition to accommodate the increasing demand for specializing training, especially in the labour market. In this regard, a recent study by the International Institute for Educational Planning (UNESCO IIEP) shows that there are fewer pathways for students to prepare for transitioning to the labour market than for entering and progressing through higher education\footnote{\url{https://unesdoc.unesco.org/ark:/48223/pf0000381545/PDF/381545eng.pdf.multi}}.

\subsection{Learning}
\paragraph{}
Among the various aspects that comprise the education field, learning is the one that requires the most careful treatment. The inherently complex domain lends itself to a wide range of forms and means, each with its techniques and quirks. Learning may happen at home, at work, or even on the go; thus, it is not limited to the classroom only nor confined to formal settings in general. Learning activities can (and should be) tailored around the individual. This personalization process is critical when targeting neurodiverse\footnote{\url{https://en.wikipedia.org/wiki/Neurodiversity}} profiles or students with accessibility needs. Not only is content's form fundamental, but delivery and additional aids are required to make a learning experience impactful.

\paragraph{}
\textbf{Educational resources} are the primary means to help (and sometimes assess) students in their learning journey. They can be of various kinds such as videos, interactive tutorials, pdf texts, images, podcasts and many others. Platforms such as \textit{Google Classroom\footnote{\url{https://edu.google.com/workspace-for-education/classroom/}}}, \textit{Microsoft Teams\footnote{\url{https://www.microsoft.com/en-us/microsoft-teams/group-chat-software}}}, or custom \textit{Moodle\footnote{\url{https://moodle.org/}}} deployments are common ways to deliver content in formal educational settings, but \textit{YouTube\footnote{\url{https://www.youtube.com/}}}, \textit{Wikipedia\footnote{\url{https://en.wikipedia.org/wiki/Wikipedia}}}, and platforms like \textit{Udemy\footnote{\url{https://www.udemy.com/}}} can reach a broader audience in informal environments. Furthermore, the increasing availability of Open Educational Resources (OER)\footnote{Open Educational Resources are public domain or open licensed educational resources \url{https://www.unesco.org/en/communication-information/open-solutions/open-educational-resources}} can help reduce teachers' material preparation and lower the accessibility barrier in terms of cost and material kind.

\paragraph{}
Learning is also crucial in the industry. The continuous technological disruptures are creating job positions in brand new fields\footnote{\url{https://www.linkedin.com/business/talent/blog/talent-strategy/linkedin-most-in-demand-hard-and-soft-skills}}. Entirely new hard skills are required to fit the openings. Moreover, soft skills are among the most sought-after skills because of their lower trainability and a slower changing pace\footnote{\url{https://www.linkedin.com/business/talent/blog/talent-acquisition/why-shell-pushes-hard-on-soft-skills}}. However, the offer seldom matches the demand. This mismatch led to the need for \textbf{upskilling} and \textbf{reskilling}. The former means teaching employees new, advanced and valuable skills to match the profile required for the next step in their current career path. The latter, instead, targets employees with a profile similar to the one required. It consists in teaching them adjacent skills and training them for their new, different job. Both of them are fundamental learning activities that take place in varied (and often dissimilar) environments.

\paragraph{}
The design of a learning experience should take into account all of these different factors. On the other hand, \textbf{a design framework should not make assumptions about content type, form, delivery, and validation while still removing any obstacle between the teacher, the student, the environment, and the learning experience.}

\subsection{Teaching}
\paragraph{}
From a student's perspective, effective teaching means 1-1 coaching. It allows teachers to target specific misunderstandings and necessities with real-time feedback and explanations relevant to the student's experiences. On the other end of the spectrum, the most teacher-effective approach is the one-to-many lecture, where the teacher prepares the material upfront for being presented to a wide audience. One of this approach's main downsides is encouraging passive learning. A study by K.R. Koedinger et al. shows that the "Doer Effect" is a causal association between practice and learning outcomes and that practising is six times more effective than reading \cite{DOER_EFFECT}. Another significant drawback lies in the motivational aspect. \textbf{Active learning} is more effective in learning outcomes and motivation than passive learning \cite{ACTIVE_LEARNING}. Despite that, a recent article on PNAS by L. Deslauriers et al. highlights a negative correlation between actual learning and the feeling of learning in the students \cite{ACTIVE_LEARNING_2}. However, \textbf{gamification} and serious games gained consensus as tools to motivate people to engage in beneficial activities, even if seen as unrewarding or tedious \cite{GAMIFIED_APPLICATIONS}. 

\paragraph{}
Individual coaching is rarely feasible due to poor scalability, whilst one-to-many general training is scalable but lacks individualization altogether. Adaptive learning combines the benefits of individualized delivery and manageability by leveraging software personalization, while gamification can be used to enhance motivation through personalized rewards or cooperative and competitive activities.

\section{Existing technology}
\paragraph{}
Adaptive learning technologies have gained traction over the last decade. Existing solutions have been successful in both domain-specific \cite{PRUSTY_ADAPTIVE} and institution-wise implementations. In 2015, the Colorado Technical University (CTU) reported that, following their \textit{Intellipath\texttrademark} adoption, the course pass rate rose sharply by 27\%, and the average grade and retention rate were also significantly affected. CogBooks\footnote{\url{https://www.cogbooks.com/}} conducted a pilot study with Arizona State University (ASU), resulting in findings similar to CTU.

\paragraph{}
SmartSparrow\footnote{\url{https://www.smartsparrow.com/}} (recently acquired by the industry-leading Pearson) provides a user-friendly WYSIWYG content authoring tool to create interactive online experiences. Realizeit, Cerego and CogBooks create content for their adaptive platforms by partnering with institutions like ASU, the American Psychological Association, or directly with industrial partners and customers. Each platform also provides in-depth analytics on the students' performance, errors, time spent learning, and other related metrics. Those solutions solve the problem of individualization and verticality of linear learning paths but overlook a fundamental factor: familiarity with the tools and the environment. Job candidates in software engineering, for example, are expected to be somewhat comfortable with industry-leading tools and methodologies. Therefore, not only delivering interactive exercises directly on those tools enables richer experiences but also better prepares the students for their future. High school electronics students, instead, may benefit from hands-on experience with physical devices like Arduinos\footnote{\url{https://www.arduino.cc/}}. Moreover, integrating education-ready tools like pi-top kits\footnote{pi-top produces high-quality educational kits for electronics and robotics based on the Raspberry Pi. \url{https://www.pi-top.com/}} allows access to already existing quality resources.

\paragraph{}
Novel interactive learning experiences can emerge even with the use of available technologies. Augmented and virtual reality are still emerging, but education and industry are already taking advantage of their benefits \cite{VR_AR_EDUCATION}. In the classroom, they are used to promote interactivity and generate engagement. In the industry, they found use in, among the others, remote maintenance on industrial devices, surgery training programs, aviation, and even military equipment\footnote{Since 2019 Microsoft and the US Army have collaborated on using HoloLens to enhance soldiers' situational awareness \url{https://news.microsoft.com/transform/u-s-army-to-use-hololens-technology-in-high-tech-headsets-for-soldiers/}}. Similarly, the use of voice user interfaces (like virtual assistants) may be a means to mimic a study companion that can ask questions, give feedback and help with misunderstandings. Adaptive educational technology should allow and encourage the creation of these varied activities to prepare the students for dynamic and changing environments where learning and adaptability are essential.

\section{Thesis objectives and structure}
The main focus of current implementations of adaptive learning technologies is the content personalization engine. However, although being a crucial component of adaptive experiences, the \label{delivery} delivery of said personalized content happens through a custom-designed student platform. That may be a good fit for some activities, but it is a limiting factor for others, such as those in the aforementioned examples. This document aims to provide an \textbf{open, content-agnostic, and extensible framework to enable the design and consumption of adaptive and gamified learning experiences.} To achieve that, it is necessary to define beforehand what learning pathways are, what are they composed of, and how can they be personalized (\hyperref[chapter:two]{Chapter 2}). Afterwards, in \hyperref[sec:gamification]{section 2.3}, we will address how gamification is linked to adaptivity and its crucial role in learning pathways. After an excursus on the used technologies in \hyperref[chapter:three]{chapter 3}, \hyperref[chapter:four]{chapter 4} will describe the proposed architecture, the motivations behind the critical choices from both a teacher's and a student's perspective, its implementation, and how it can be extended to include new tools for the teacher and delivery mediums for the student. Lastly, the \hyperref[chapter:five]{closing chapter} will include a glimpse into future directions and possibilities enabled by the proposed solution.

%% file: chapters/chapter_1/footnote_1.tex
Data from UNESCO \cite{UNESCO_FLP} and Statista (retrieved on 29 June 2022). \\
\url{https://www.statista.com/statistics/1226999/net-enrollment-rate-in-primary-school-worldwide/} and \url{https://www.statista.com/statistics/1227022/net-enrollment-rate-in-secondary-school-worldwide/}

%% file: chapters/chapter_2/chapter_2.tex
\chapter{Formalization}
\label{chapter:two}
\paragraph{}
A \textbf{learning path}\footnote{\url{ https://en.wikipedia.org/wiki/Learning_pathway}} is a route a learner takes through the domain of learning activities that allows them to build knowledge progressively. Learning path designers create them to help students move from \textit{an} initial competency set toward \textit{a} goal competency set via a \textit{fixed} sequence of activities. They are usually tailored to the average student; therefore, they do not fit the majority of learners by design\footnote{TED Talk by Todd Rose (former professor at Harvard and founder of the Laboratory for the Science of Individuality) \url{https://www.youtube.com/watch?v=4eBmyttcfU4}}. An effective learning path is a \textbf{student-specific} program that helps them move from \textit{their} initial competency set to \textit{their} goal competency set via a \textit{flexible} sequence of activities that considers \textbf{individual strengths and weaknesses}\footnote{\url{ https://en.wikipedia.org/wiki/Individual_Learning_Plan}}. The remainder of this chapter will describe how the generation of a suitable learning path can be formulated as a planning problem\footnote{\url{https://en.wikipedia.org/wiki/Automated_planning_and_scheduling}} in the domain of learning activities.

\section{The planning domain}
\begin{figure}[htbp]
    \centering
    \subfloat[the planning domain hypergraph]{
        \label{hypergraphA}
        \includegraphics[width=.45\textwidth]{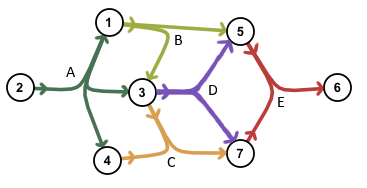}
    }
    \par
    \subfloat[admissible solution (B, D, E)]{
        \label{hypergraphB}
        \includegraphics[width=.4\textwidth]{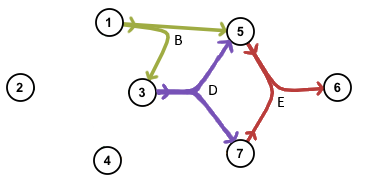}
    }
    \hfill
    \subfloat[admissible solution (A, C, E)]{
        \label{hypergraphC}
        \includegraphics[width=.4\textwidth]{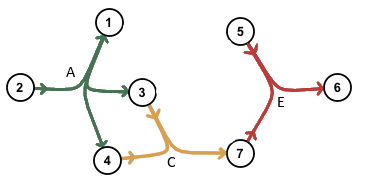}
    }
    \captionsetup{width=0.785\textwidth}
    \caption[]{\hyperref[hypergraphA]{Figure (a)} shows an example domain hypergraph. Given the initial competency set \{1, 2, 5\} and goal \{6\}, figures \hyperref[hypergraphB]{(b)} and \hyperref[hypergraphC]{(c)} show two different admissible learning paths\footnotemark.}
    \label{fig:domainHypergraph}
\end{figure}

\paragraph{}
Durand et al. propose a model to build a learning design recommendation system based on graph theory using cliques \cite{GRAPH_FORMALIZATION}. Similarly, we want to formulate a hypergraph in terms of competencies and learning activities. Although the literature does not agree on the term "\textbf{competency}", the definition "Competencies are skills, knowledge, and capabilities that individuals should have possessed when completing assigned tasks or achieving the goals", proposed by Chung \& Lo in 2007, serves our purpose. \textbf{Learning activities} are activities designed to create the conditions for learning\footnote{\url{ https://newlearningonline.com/learning-by-design/glossary/learning-activity}}. They correspond to the "when completing assigned tasks or achieving the goals" part of the competencies definition. In other words, learning activities link competencies in terms of "\textbf{prerequisites}" and "\textbf{effects}". Therefore, we see competencies as vertices and learning activities as directed hyperedges connecting the competencies required to those that should be obtained upon completing the activity. \hyperref[hypergraphA]{Figure 2.1 (a)} shows a hypergraph modelling a simple domain composed of competencies (1..7) and activities (A..E).

\footnotetext{Images based on \url{https://mathematica.stackexchange.com/a/58758}}

\noindent
A simplistic initial formulation may be:
\begin{quote}
    Student competencies are defined by the set $C_{stud} = \{c_1, c_2, ..., c_n\}$
    \par
    A learning activity $a = (a_{pre}, a_{post})$ is an action that can be taken iff $a_{pre} \subseteq C_{stud}$ and results in the new $C_{stud}$ to be $C_{stud} \cup a_{post}$
    \par
    A goal is a set of competencies $G = \{c_1, c_2, ..., c_n\}$
    \par
    A learning path is a sequence of learning activities $S = \{a_1, a_2, ..., a_n\}$ such that, given a goal $G$
    \[G \subseteq C_{stud} \cup \bigcup_{j=1}^{n} a_{jpost} \land \forall i \in [1..n] : a_{ipre} \in C_{stud} \cup \bigcup_{j=1}^{i-1} a_{jpost}\]
\end{quote}
\paragraph{}
Although correct, this first definition exemplified in \hyperref[fig:domainHypergraph]{figure 2.1} does not consider any other factor than the domain itself (that only includes competencies and activities). Furthermore, the planning domain may evolve and even change, the student's behaviour is dynamic and partially observable, and the result of the student's actions is uncertain. Planning should also consider that the student may alternate different activities or acquire competencies in other ways. Because learning activities aim to create the conditions for learning, and those conditions are different for each individual, their effect is not even guaranteed.

\paragraph{}
Instead, following the approach proposed by Bucchiarone et al. \cite{DYMANIC_FRAGMENT_PLANNING}, planning in a dynamic system relies on a learning model (or context model) composed of \textbf{learning properties}, each describing an aspect of the system (e.g. student competencies, ...). A \textbf{learning context} captures the status of all learning properties. The status of each learning property can evolve naturally through \textbf{learning fragments} or unpredictably through external changes (e.g. obtaining a certificate or an award).

\section{Learning fragments}
\label{sec:learningFragments}
\paragraph{}
Learning fragments are a structured composition of learning activities. They are a more realistic representation of an educational process that can include complex flows of other activities. Fragments behave like regular learning activities: they have prerequisites which limit their applicability and effects that can influence the learning context. The following sections will further detail fragments, their creation, and their application with the help of a motivating scenario.

\subsection{Motivating scenario}
\label{motivatingScenarioSection}
\paragraph{}
The chosen scenario is part of an introductory statistics course that fits a broader data science program. Its goal is for the student to understand the concepts of average, median, and the difference between them. In other words, the students should have obtained those competencies when completing the fragment.

\paragraph{}
We have identified four kinds of concrete learning activities needed to define the scenario:
\begin{itemize}
    \item lessons
    \item close-ended questions
    \item quizzes
    \item coding activities
\end{itemize}

\paragraph{}
Each concrete activity serves a different educational purpose. Lessons are activities that should augment the student's knowledge and understanding of a topic. Closed-ended questions are short assignments to get students to apply the concepts learned in lessons immediately and gauge their initial understanding. Quizzes follow a review lesson and verify if the student is ready to move to the next topic. Lastly, coding activities occur at the end of the fragment to solidify theoretical knowledge and gain hands-on experience on the topic.

\begin{figure}[htbp]
    \centering
    \includegraphics[width=\textwidth]{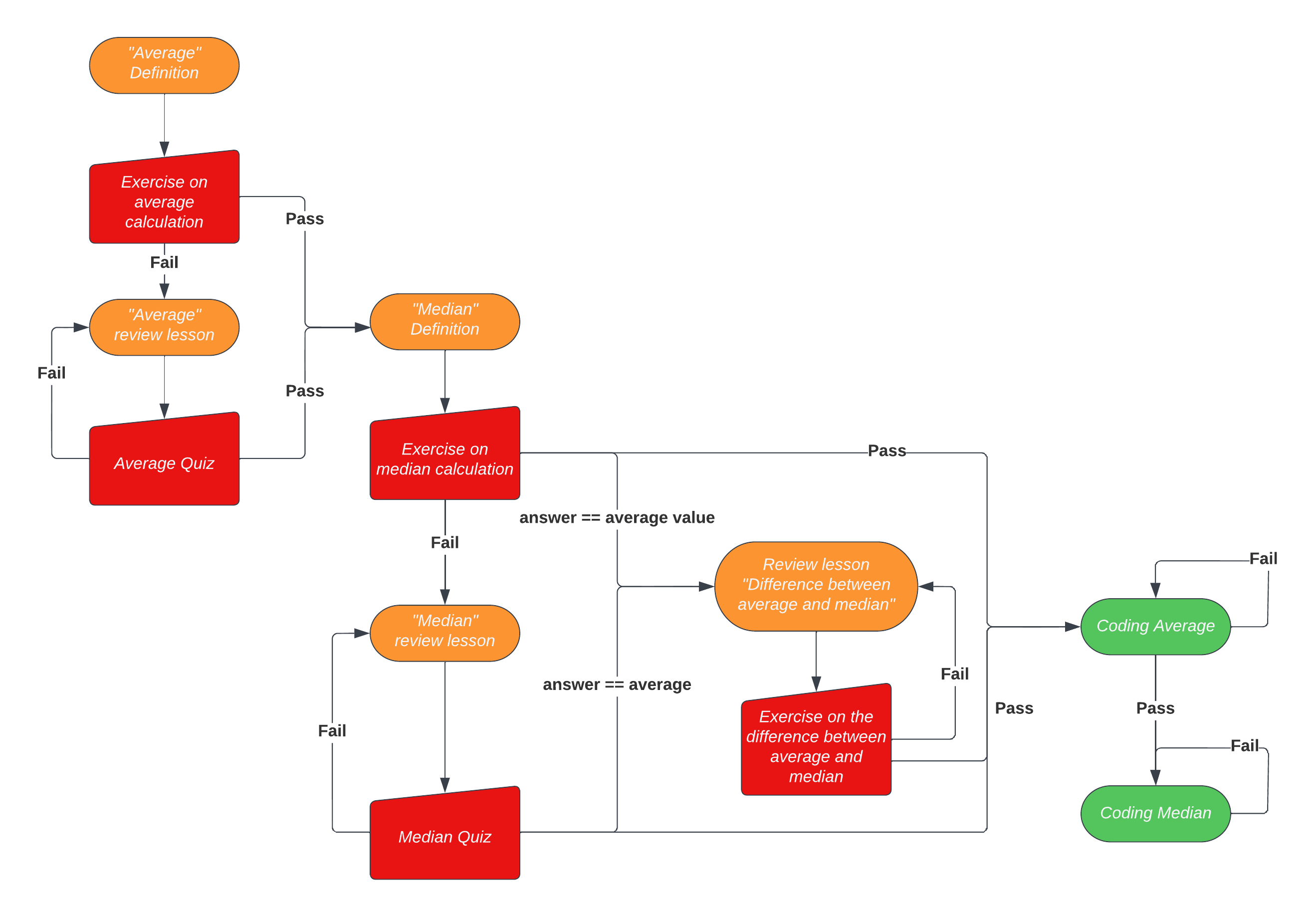}
    \caption{A learning fragment for an introductory statistics course}
    \label{fig:motivatingScenario}
\end{figure}

\paragraph{}
We want students to first learn about the average, then the median, and then how to code them. We also want students to \textit{understand} the average, the median, and the difference between them, so the fragment includes a review lesson and an additional quiz in case the student fails the close-ended question. Once the student has completed the "average" section, they move to the median, where they will face the same pattern for recovery used for the average. Suppose that the closed-ended question requires the student to calculate the median. Since understanding the difference is part of the goal, if the student answers with the average value instead of the median, they should be given a particular review lesson that highlights the differences instead of a regular review lesson. All those different alternative paths merge before the coding activities. While the other exercises have a known correct answer, coding activities do not. Furthermore, to be considered correct, a program may not only be required to have a correct output (e.g. through unit tests) but also to have some properties on the source code itself (e.g. low cyclomatic complexity\footnote{\url{ https://en.wikipedia.org/wiki/Cyclomatic_complexity}} or proper encapsulation\footnote{\url{https://en.wikipedia.org/wiki/Encapsulation_(computer_programming)}}) or on its execution (e.g. time or memory constraints). For the sake of simplicity, the transitions in this scenario's coding activity only include "pass" and "fail" without further specifying the criteria. However, as evidenced by the previous example, fragments must not pose any limit on the relationship between inner activities. The complete fragment diagram appears in \hyperref[fig:motivatingScenario]{figure 2.2}.

\subsection{Towards more complete learning fragments}
\paragraph{}
As described in \hyperref[motivatingScenarioSection]{section 2.2.1}, fragments are self-contained structured activities, but sometimes educators do not need to be specific when assigning exercises or activities. Suppose a teacher wants students to train on self-balancing binary search trees to cement their knowledge better. Any chosen exercise (or structured activity/learning fragment) would fit the task as long as they satisfy specific properties, such as being about the assigned topic and being in the same difficulty category. This scenario leads to the need to be able to define activities abstractly.

\subsubsection{Abstract Learning Activities}
\paragraph{}
"An abstract learning activity is defined at design time in terms of the goal it needs to achieve, expressed as learning context configurations to be reached, and is automatically \textbf{refined at run time} into an executable learning path, considering the set of available fragments, the current learning context, and the learning goal to be reached" \cite{DYMANIC_FRAGMENT_PLANNING}.

\paragraph{}
The refinement process aims to automatically compose available fragments to fulfil the goal of the abstract activity, given the current context configuration. Abstract activities may also contain other abstract activities, so further refinements may be required.

\paragraph{}
This approach's advantage is twofold: thanks to runtime refinement, teachers can spend less time focusing on irrelevant details, while students can benefit from a more accurate personalization. During the refinement process, abstract activities can be replaced with the best fragment according to the most up-to-date student profile, his understanding of the topic, and available educational resources. For example, students with attention deficit hyperactivity disorder (ADHD) may benefit from consuming fragments composed of bite-sized activities instead of activities that require long-lasting concentration so they can take more frequent breaks\footnote{\url{https://www.verywellmind.com/help-for-students-with-adhd-20538}} \cite{ADHD_EDUCATION}. Similarly, those who have shown more significant gains when exposed to video resources (i.e. instead of textual resources) may benefit from being suggested videos more often than others.

\paragraph{}
Furthermore, runtime refinement facilitates the reuse and enhances the updatbility of existing content. Fragments can be as specific as the one proposed in the scenario and as complex and abstract as a course syllabus or even an entire academic degree programme. A course outline may still be valid and relevant, even when educational resources are being updated, new fragments are added, or existing ones are changed. Likewise, specific fragments and activities can be reused (if still appropriate) when the course outline is modified, i.e. when topics are added/removed, or their order is changed.

\subsection{Formal framework}

\newcommand{\tuple}[1]{\ensuremath{\langle #1 \rangle}}
\newcommand{\CPDiagram}{\ensuremath{c}}
\newcommand{\CPDiagrams}{\ensuremath{C}}
\newcommand{\CPDLoc}{\ensuremath{L}}
\newcommand{\CPDLocInit}{\ensuremath{\CPDLoc_0}}
\newcommand{\CPDEvent}{\ensuremath{e}}
\newcommand{\CPDEvents}{\ensuremath{E}}
\newcommand{\CPDTrans}{\ensuremath{T}}
\newcommand{\SAnnotation}{\ensuremath{\textit{Ann}}}
\newcommand{\Current}{\ensuremath{curr}}
\newcommand{\Refinements}{\ensuremath{R}}
\newcommand{\SAnnEffects}{\ensuremath{\textit{Eff}}}
\newcommand{\SAnnPreconditions}{\ensuremath{\textit{Pre}}}
\newcommand{\SAnnGoals}{\ensuremath{\textit{Goal}}}
\newcommand{\SCompensations}{\ensuremath{\textit{Comp}}}
\newcommand{\Process}{\ensuremath{l}}
\newcommand{\Execution}{\ensuremath{E}}
\newcommand{\System}{\ensuremath{\mathcal{S}}}
\newcommand{\SysExec}{\ensuremath{\Lambda}}
\newcommand{\History}{\ensuremath{h}}
\newcommand{\Executions}{\ensuremath{\mathcal{E}}}
\newcommand{\ApplicationModels}{\ensuremath{M}}
\newcommand{\Fragments}{\ensuremath{\mathcal{F}}}
\newcommand{\CFragments}{\ensuremath{F_C}}
\newcommand{\RFragments}{\ensuremath{F_R}}
\newcommand{\Fragment}{\ensuremath{f}}
\newcommand{\Service}{\ensuremath{s}}
\newcommand{\Services}{\ensuremath{SSS}}
\newcommand{\SStates}{\ensuremath{S}}
\newcommand{\SAction}{\ensuremath{a}}
\newcommand{\SActions}{\ensuremath{A}}
\newcommand{\SInputs}{\ensuremath{A_{in}}}
\newcommand{\SOutputs}{\ensuremath{A_{out}}}
\newcommand{\SAbstracts}{\ensuremath{A_{abs}}}
\newcommand{\SConcretes}{\ensuremath{A_{con}}}
\newcommand{\SIStates}{\ensuremath{\SStates_0}}
\newcommand{\STrans}{\ensuremath{T}}
\newcommand{\SEffect}{\ensuremath{E}}
\newcommand{\SPrecondition}{\ensuremath{P}}
\newcommand{\SFaults}{\ensuremath{O'}}
\newcommand{\SDecisions}{\ensuremath{DP}}
\newcommand{\SDecision}{\ensuremath{D}}
\newcommand{\AS}{\ensuremath{\xi}}
\newcommand{\ASInit}{\ensuremath{\mathcal{I}}}
\newcommand{\ASFinal}{\ensuremath{\mathcal{G}}}
\newcommand{\Evolution}{\ensuremath{\mathcal{E}}}
\newcommand{\GSystem}{\ensuremath{\Gamma}}
\newcommand{\STS}{\ensuremath{\Sigma}}

\subsubsection{Learning Properties}
\paragraph{}
\noindent
Every learning property is modelled with a \textit{learning property diagram}: a state-transition system capturing all possible property values and value changes, each labeled with the corresponding event.

\paragraph{}
\noindent
A learning property diagram is a tuple $\CPDiagram = \tuple{\CPDLoc, \CPDLocInit, \CPDEvents, \CPDTrans}$, where:
\begin{itemize}
    \item $\CPDLoc$ is a set of learning states and $\CPDLocInit\subseteq\CPDLoc$ is a set of initial states;
    \item $\CPDEvents = \CPDEvents_{unc}\cup\CPDEvents_{cnt}$ is a set of learning events, where $\CPDEvents_{unc}$ is a set of uncontrollable and $\CPDEvents_{cnt}$ is a set of controllable events, such that $\CPDEvents_{cnt}\cap\CPDEvents_{unc}=\emptyset$; 
    \item $\CPDTrans\subseteq\CPDLoc\times\CPDEvents\times\CPDLoc$ is a transition relation.
\end{itemize}
$\CPDLoc(\CPDiagram)$, $\CPDEvents(\CPDiagram)$, etc. indicate the corresponding elements of $\CPDiagram$.

\paragraph{}
\noindent
A learning context $\CPDiagrams$ is defined as a set of individual learning property diagrams; thus, a space of \textit{learning context configurations} is defined as $\CPDLoc=\prod_{\CPDiagram \in \CPDiagrams}\CPDLoc(c)$ and the set of controllable events is  $\SEffect_{cnt}=\bigcup_{\CPDiagram \in \CPDiagrams} \CPDEvents_{cnt}(\CPDiagram)$ .

\subsubsection{Learning Fragments}
\paragraph{}
\noindent
A learning fragment is a tuple $\Process = \tuple{\SStates, \SIStates, \SActions, \STrans, \SAnnotation}$, where:
\begin{itemize}
    \item $\SStates$ is a set of states and $\SIStates\subseteq\SStates$ is a set of initial states;
    \item $\SActions=\SConcretes\cup\SAbstracts$ is a set of activities, where $\SConcretes$ is a set of concrete activities and $\SAbstracts$ is a set of abstract activities. $\SConcretes$, and $\SAbstracts$ are disjoint sets;
    \item $\STrans\subseteq\SStates\times\SActions\times\SStates$ is a transition relation;
    \item $\SAnnotation=\tuple{\SAnnPreconditions, \SAnnEffects, \SAnnGoals}$ is a process annotation, where $\SAnnPreconditions$ is the precondition labelling function that links $\SConcretes$ to the learning properties, $\SAnnEffects$ is the effect labelling function that links $\SConcretes$ to the controllable events that contribute to mutating the learning properties, and $\SAnnGoals$ is the goal labelling function that links $\SAbstracts$ with its goal state.
\end{itemize}

\subsubsection{Learning Configuration}
\paragraph{}
\noindent
A learning configuration is a tuple $\SysExec = \tuple{\ASInit,\Fragments}$ where:
\begin{itemize}
    \item $\ASInit\in\CPDLoc(\CPDiagram_1)\times\ldots\times\CPDLoc(\CPDiagram_n), \CPDiagram_i\in\CPDiagrams$ is the current configuration of learning property diagrams;
    \item $\Fragments$ is the set of available learning fragments.
\end{itemize}
$\ASInit(\SysExec)$, $\Fragments(\SysExec)$ indicate the corresponding elements of a configuration $\SysExec$.

\subsubsection{Goal}
\paragraph{}
\noindent
Goals are not simply states $l \in \CPDLoc$ to be reached (“reachability goals”), but also conditions on the whole plan execution paths (“extended goals"). A goal $\ASFinal$ is satisfied when the current learning configuration $\SysExec$ is such that $\ASInit(\SysExec) \models \ASFinal$ and the execution plan used to reach the configuration satisfies the extended conditions \cite{GOAL_DEFINITION}.

\subsubsection{Refinement}
\paragraph{}
\noindent
An adaptation problem is generically defined as a tuple $\AS = \tuple{\SysExec, \ASFinal}$, where $\SysExec$ is the current learning configuration and $\ASFinal$ is an adaptation goal over $\CPDiagrams$.

\paragraph{}
\noindent
$\SysExec(\AS)$ and $\ASFinal(\AS)$ indicate the corresponding elements of a problem $\AS$.

\paragraph{}
\noindent
Refinement is a particular kind of adaptation mechanism used when $\SAction$ is an abstract activity. It means solving an adaptation problem $\AS$ where $\ASFinal(\AS) = \SAnnGoals(\SAction)$. $\SysExec'$ is obtained from $\SysExec$ using the new refined fragment.

\subsubsection{Planning}

\begin{figure}[htbp]
    \centering
    \includegraphics[width=\textwidth]{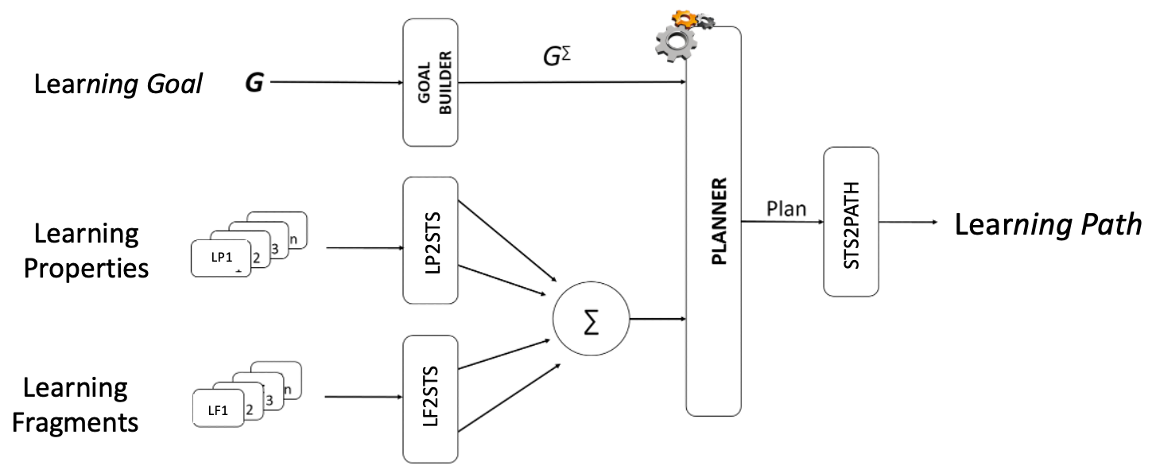}
    \caption{The planning process needs a goal and the learning properties and fragments in their STS form}
    \label{fig:planningSchema}
\end{figure}

\paragraph{}
\noindent
A planning domain is derived from adaptation problem $\AS$ that aims to achieve an adaptation goal. In particular, a set of $n$ learning fragments  ($f_1,\dots,f_n\in\Fragments(\SysExec)$) and a set of $m$ learning property diagrams ($c_1,\dots,c_m\in\CPDiagrams$) are transformed into state transition systems (STS) using transformation rules similar to those presented in \cite{GOAL_DEFINITION}. The planning domain $\STS$ is obtained as the product of the STSs $\STS_{f_1}$ \ldots $\STS_{f_n}$ and $\STS_{c_1}$ \ldots $\STS_{c_m}$, where STSs for fragments are synchronized on preconditions and effects.

$$
\STS =
\STS_{\Process_1}\|\dots\|\STS_{\Process_n}~\|~\STS_{c_1}\|\dots\|\STS_{c_m}
$$

\noindent
The initial state $r$ of the planning domain is derived from the current configuration $\ASInit(\SysExec)$, by interpreting it as states of the STSs defining the planning domain. Similarly, the learning goal $\ASFinal(\AS)$ is transformed into a planning goal $\rho$ by interpreting the configurations in $\ASFinal(\AS)$ as states in the planning domain. Once everything has been transformed, the planner generates a plan $\STS_c$ that needs to be transformed back into its executable form. The process is illustrated in \hyperref[fig:planningSchema]{figure 2.3} 

\paragraph{}
\noindent
For further reading on the topic, see "Dynamic Adaptation of Fragment-based and Context-aware Business Processes" and "Incremental Composition for Adaptive by-Design Service Based Systems" by Bucchiarone et al. \cite{DYMANIC_FRAGMENT_PLANNING}\cite{DYMANIC_FRAGMENT_PLANNING_2}.

\section{Gamified learning paths and activities}
\label{sec:gamification}
\paragraph{}
Gamification is the strategic attempt to enhance systems, services, organizations, and activities by creating similar experiences to those experienced when playing games in order to motivate and engage users\footnote{\url{https://en.wikipedia.org/wiki/Gamification}}. Gamified applications are increasingly popular: Duolingo\footnote{\url{https://www.duolingo.com/}} uses gamification to teach languages, and Apple Fitness\footnote{\url{https://en.wikipedia.org/wiki/Fitness_(Apple)}} encourages its users to "close" the daily activity rings by including badges, challenges, and competitions. Minecraft Education\footnote{\url{https://education.minecraft.net/}} takes the opposite approach, with the base game being enriched with educational tools to exploit its popularity (with significant results \cite{MINECRAFT_EDU}). Location-based games have been used to promote personal hygiene and social distancing to limit the COVID-19 spreading \cite{MEET_DURIAN}. In the domain of software engineering, POLYGLOT explores the application of gamification mechanisms to mixed modelling and programming exercises \cite{POLYGLOT}.

\paragraph{}
Designing a gamified tool is not only adding points and badges. It means creating a game narrative that guides players through increasingly complex challenges, keeping them engaged with social activities such as group work or competitions. It means providing immediate feedback (as expected from a game-like environment) and students taking autonomous choices to progress down the individually decided path. Gamification is not an add-on. Instead, gamification mechanisms are fundamental to the learning path personalization process in two ways. Not only they do keep the students engaged, but they can also be used as tools to gain insight into the student's behaviour from a different perspective and thus help generate a more personalized and engaging learning path.

\paragraph{}
We define \textbf{gamification properties} (such as points or badges), \textbf{gamification fragments}, and \textbf{engagement configuration} $\GSystem$ analogously to the corresponding learning properties, learning fragments, and learning configuration. We can therefore use the same planning technique to generate the engagement path associated with a specific learning path. 

\paragraph{}
A system $\System$ can then be defined as a tuple $\System = \tuple{\SysExec,\GSystem}$ and its evolution, which involves both the learning configuration and the engagement configuration, is a function $\Evolution : \tuple{\SysExec,\GSystem}\rightarrow\tuple{\SysExec',\GSystem'}$. Those two subsystems are not independent: gamification mechanisms influence the evolution of the learning system (e.g. by "unlocking" an activity when a certain point threshold is reached) and the learning configuration influence the adaptation of gamification mechanisms (i.e. the required number of points necessary to "unlock" the activity). This is necessary because, in order to increase engagement, the gamification mechanisms must be \textbf{calibrated} according to the underlying activities.

\paragraph{}
If we were to plot engagement versus learning effectiveness, we would aim to keep the students in the upper right corner of the plot. Both gamification mechanisms and learning activites influence both metrics. No matter what gamification mechanisms are in place, if a student fails several exercises in a row and does not see any chance of success or no trace of progression, they would surely get demotivated. Likewise, if gamification mechanisms were to reward easy exercises too much, the student would still have a reasonably high engagement, but their learning quality would surely diminish by not engaging in challenging activities. By influencing each other during their evolution, the learning system and the engagement system would be able to maintain the correct balance.

\begin{figure}[htbp]
    \centering
    \subfloat[fragment with abstract activity]{
        \label{fig:gamificationA}
        \includegraphics[width=.75\textwidth]{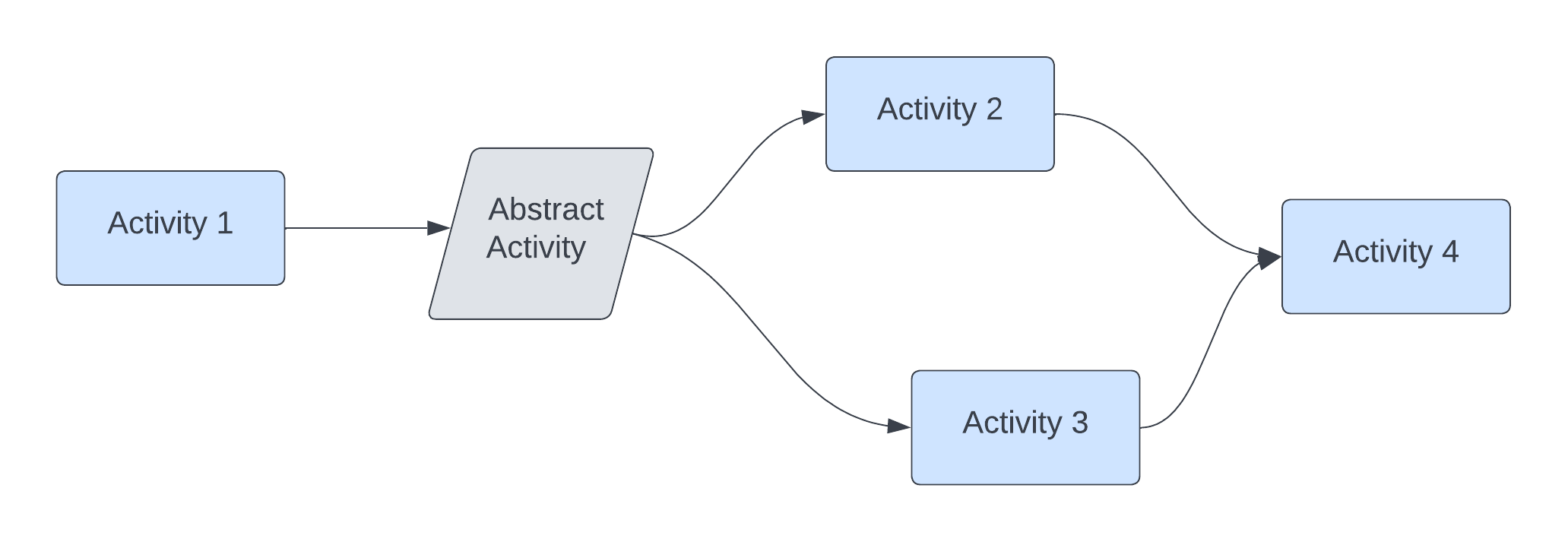}
    }
    \par
    \subfloat[easy refinement]{
        \label{fig:gamificationB}
        \includegraphics[width=.75\textwidth]{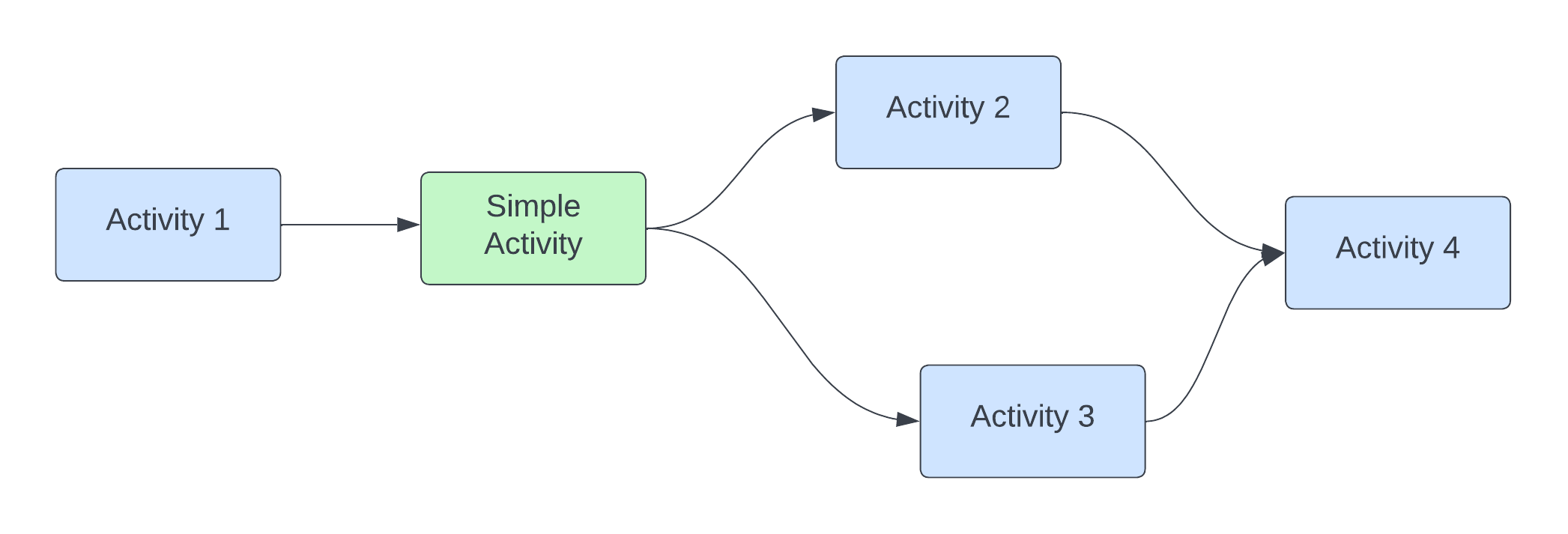}
    }
    \par
    \subfloat[hard refinement]{
        \label{fig:gamificationC}
        \includegraphics[width=.75\textwidth]{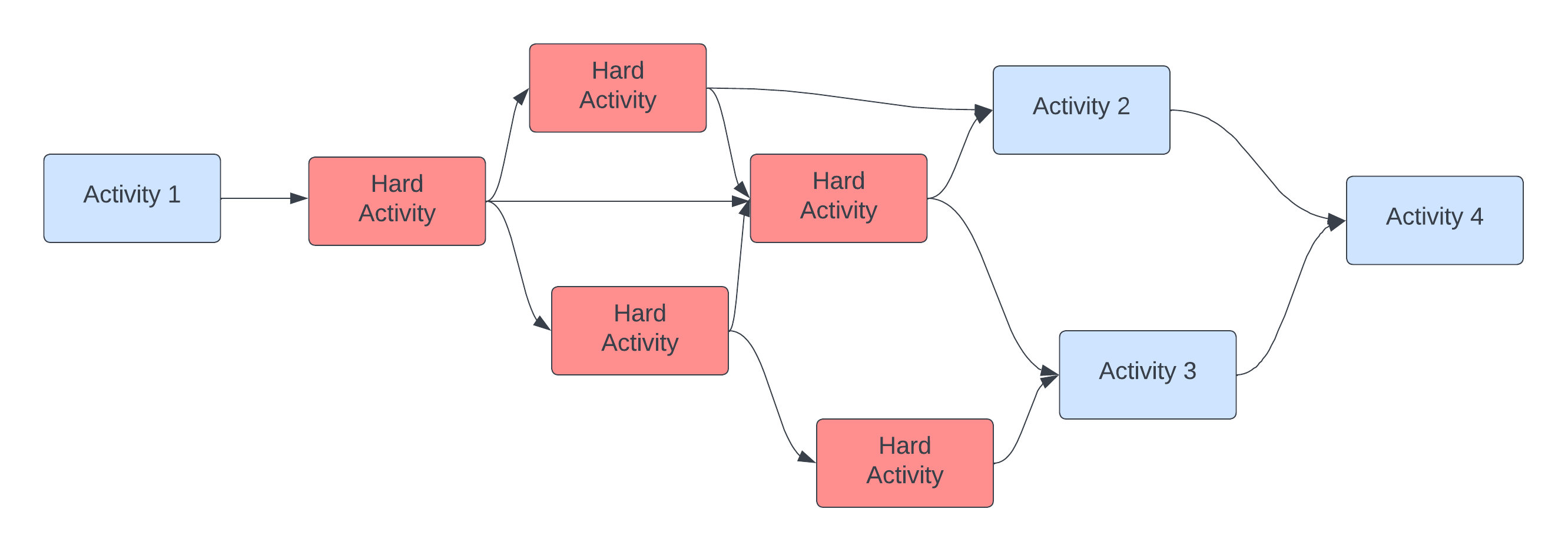}
    }
    \captionsetup{width=0.785\textwidth}
    \caption[]{Different results of the runtime refinement}
    \label{fig:gamificationGraph}
\end{figure}

For example, given the fragment shown in \hyperref[fig:gamificationA]{figure 2.4 (a)}, the runtime refinement phase could produce a result similar to \hyperref[fig:gamificationB]{2.4 (b)} or \hyperref[fig:gamificationC]{2.4 (c)}, depending on the learning context configuration (i.e. the student's profile). If we wanted to promote collaboration amongst peers, a reward for engaging in an optional collaborative activity might be a reasonable solution. However, the reward would need to be adjusted based on the number of mandatory and optional collaborative activities and their duration and difficulty. Similarly, to encourage a more effective learning process, if the student were to engage only in a narrow range of activities (e.g. quizzes), he should be prompted with choices that make appear the other options more appealing than their current path. The goal is to guide the student in their active choice to engage in more challenging and fruitful activities instead of making them mandatory and lowering their engagement.

\section{Summary}
\paragraph{}
A learning path aims to guide the student in their journey from their initial competency set towards their goal through learning activities. Each activity has a set of prerequisites and effects. Fragments are complex structured activities. They can be designed for specific purposes such as clearing misunderstandings, challenging the best students, and addressing neurodiversity and accessibility needs. During the planning phase, fragments are composed according to the student's characteristics (learning context) to satisfy a given goal. During runtime refinement, abstract activities are replaced with available fragments to fulfil their design-time goal. Deeply integrated gamification mechanisms enhance engagement and give guidance for a more suitable personalization.

%% file: chapters/chapter_3/chapter_3.tex
\chapter{Technology}
\label{chapter:three}
\section{POLYGLOT}
\paragraph{}
POLYGLOT is a research project on adaptive and gamified education led by Dr Bucchiarone at Fondazione Bruno Kessler. Initial work on mixed modelling and programming exercises \cite{POLYGLOT} has been part of the MODELS 2021 conference\footnote{\url{https://conf.researchr.org/details/models-2021/models-2021-technical-papers/30/POLYGLOT-for-Gamified-Education-Mixing-Modelling-and-Programming-Exercises}} and BETT 2022\footnote{\url{https://uk.bettshow.com/}}. Despite the initial effort in the software engineering domain, the project aims to encompass education in a broader sense and create a common ground to build novel educational experiences. To achieve that, current work on POLYGLOT comprehends three main areas:

\begin{itemize}
    \item Generalization of learning activities and metamodelling.
    \item Student and teacher usability, including the definition of the learning path and different mediums to consume it.
    \item Adaptivity and feedback to address individual needs.
\end{itemize}

\paragraph{}
The initial prototype was heavily notebook-centric and focused on programming with \CSharp. It aimed to create exercises that teach students basic programming concepts (like classes) through incremental small steps guided by timely feedback (a demo video is available at \url{https://www.youtube.com/watch?v=sxsHy2PbH8E}). The validation of student submissions not only used runtime tests but also included custom Roslyn\footnote{Roslyn is the .NET Compiler Platform and provides useful APIs to build custom code analysis tools \url{https://github.com/dotnet/roslyn}} static analyzers. The project then evolved and began to integrate other languages. In particular, we ported SysML v2\footnote{\url{ https://www.omgsysml.org/SysML-2.htm}} to .NET Interactive (more about the porting in the next section) and included it in incremental exercises formed by a design and an implementation phase in the same notebook. All prototypes have gamification elements powered by FBK's Gamification Engine\footnote{\url{https://github.com/smartcommunitylab/smartcampus.gamification}}, a rule-based gamification platform that enables the definition of Drools\footnote{\url{https://www.drools.org/}} rules governing the evolution of the game elements.

\paragraph{}

\begin{figure}[htbp]
    \centering
    \includegraphics[width=\textwidth]{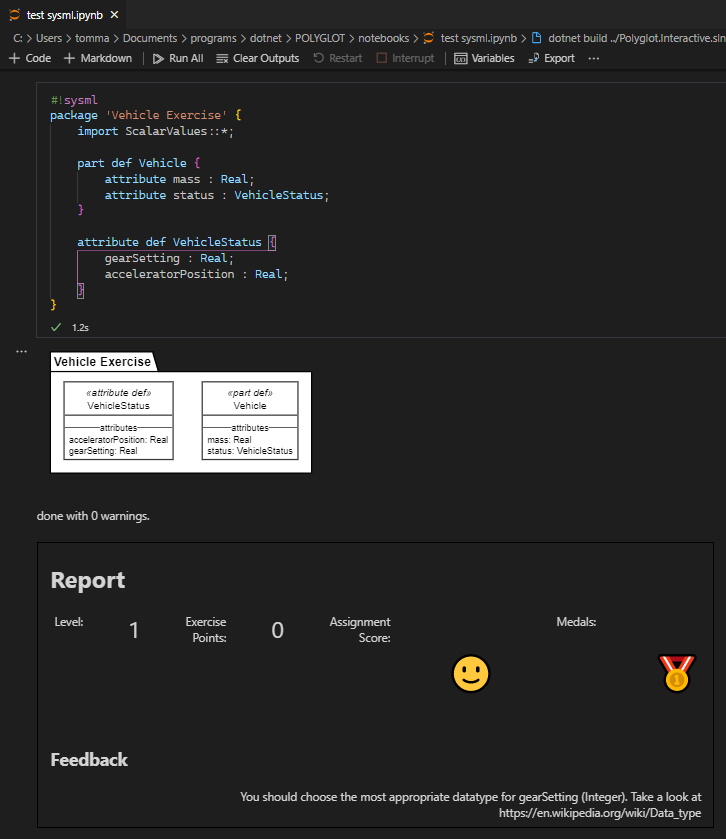}
    \caption[]{SysML v2 exercise with rich output, gamification, and feedback}
    \label{fig:sysml}
\end{figure}

Additional information on POLYGLOT and its development can be found on GitHub at \url{https://github.com/polyglot-edu}.

\section{.NET Interactive}
\paragraph{}
.NET interactive\footnote{\url{https://github.com/dotnet/interactive}} is a group of CLI tools and APIs that enable users to create interactive experiences\footnote{\url{https://devblogs.microsoft.com/dotnet/net-interactive-is-here-net-notebooks-preview-2/}}. It provides several essential functionalities for building learning experiences, such as:
\begin{itemize}
    \item \textbf{Rich output} support with MIME types and formatters\footnote{\url{https://github.com/dotnet/interactive/blob/main/docs/formatting.md}}.
    \item Multi-language support and interoperability (e.g. via variable sharing \footnote{\url{https://github.com/dotnet/interactive/blob/main/docs/variable-sharing.md}}).
    \item Loading NuGet packages at runtime.
\end{itemize}

\begin{figure}[htbp]
    \centering
    \includegraphics[width=\textwidth]{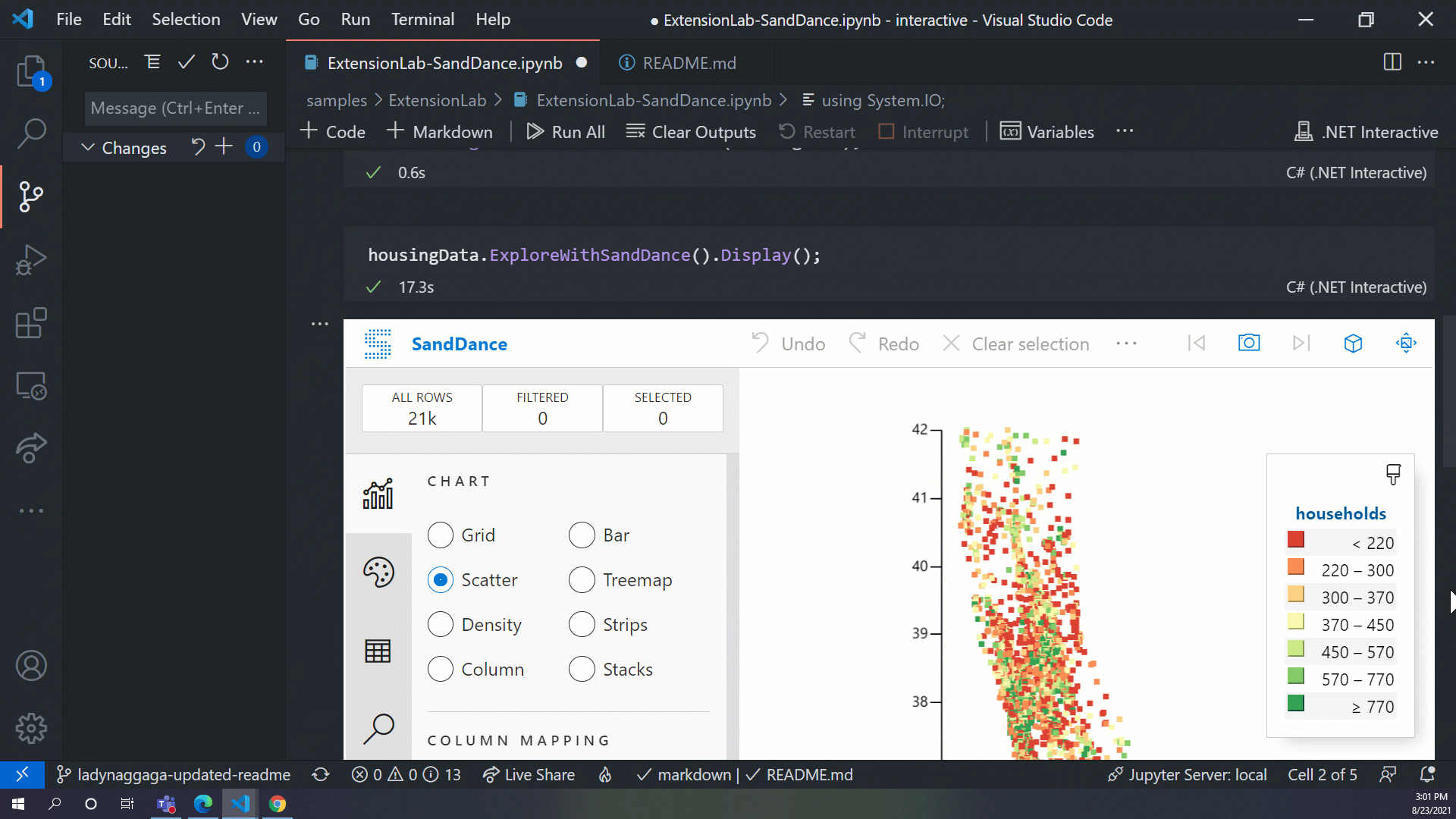}
    \caption[]{.NET Interactive rich and interactive output}
    \label{fig:richOutput}
\end{figure}

\paragraph{}
In general, the .NET Interactive execution model revolves around the concept of kernels: an abstraction of a component that accepts commands and produces output. Existing ones support several programming languages (e.g. \CSharp, \FSharp, and JavaScript), query languages (e.g. SQL), markup languages (e.g. Markdown and HTML), and others are being explored (e.g. Python). However, kernels are not limited to programming. For example, an NLP\footnote{\url{https://en.wikipedia.org/wiki/Natural_language_processing}} kernel could evaluate student solutions, answer their questions, and provide additional insights via images, audio, video, or interactive content (\hyperref[fig:richOutput]{Figure 3.2}).

\begin{figure}[htbp]
    \centering
    \includegraphics[width=\textwidth]{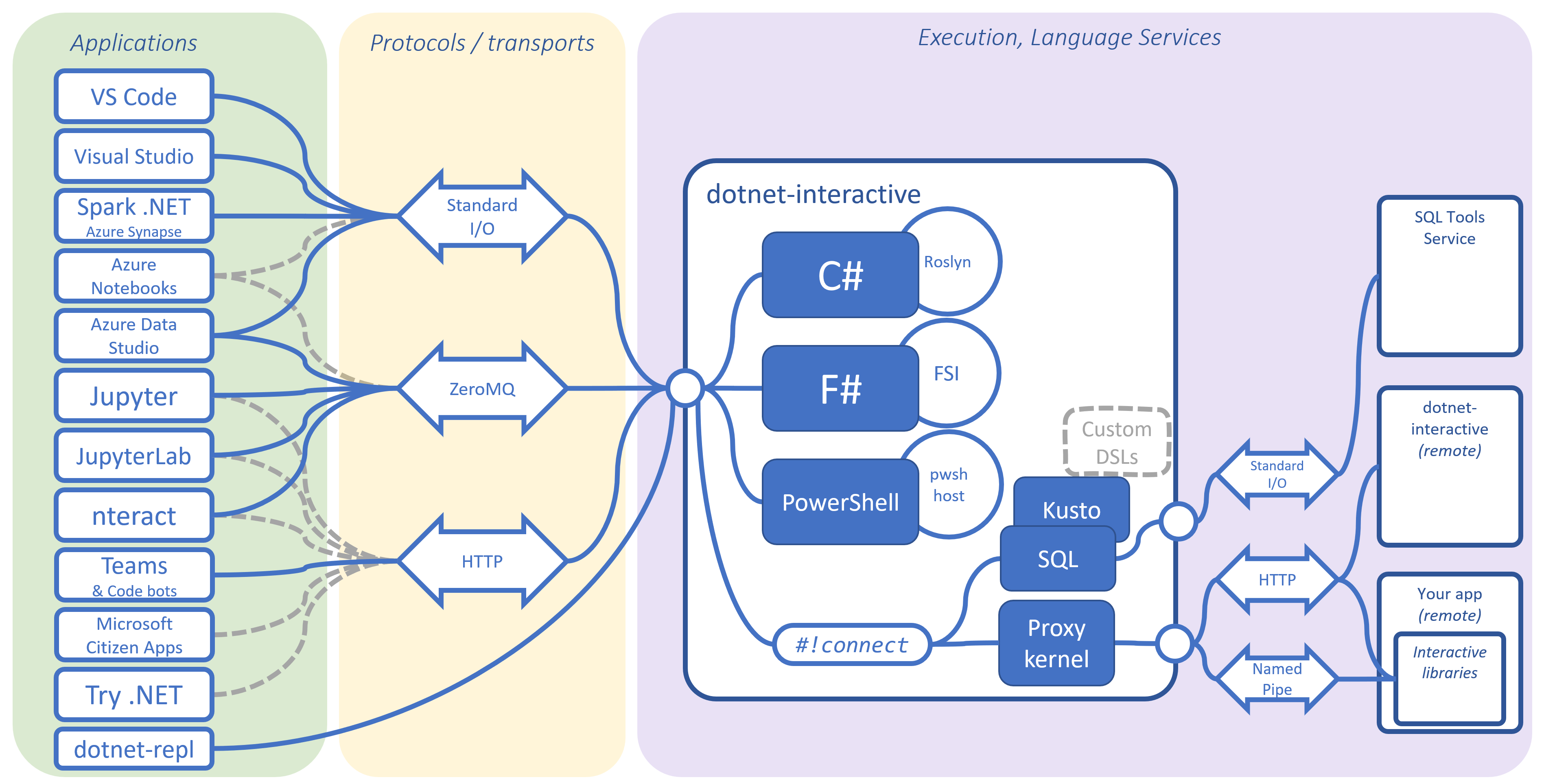}
    \caption[]{.NET Interactive architecture overview diagram\footnotemark}
    \label{fig:dotnetInteractiveArchitecture}
\end{figure}
\footnotetext{\url{https://github.com/dotnet/interactive/blob/main/docs/kernels-overview.md}\label{foot:dotnetKernelOverview}}

\paragraph{}
.NET Interactive's command-event\footnote{\url{https://en.wikipedia.org/wiki/Command_pattern}} modular architecture (see \hyperref[fig:dotnetInteractiveArchitecture]{figure 3.3}) is perfectly suitable to support our extensibility needs in several different ways:
\begin{itemize}
    \item We have successfully managed to port SysML v2 to .NET Interactive by wrapping its official kernel\footnote{\url{https://github.com/Systems-Modeling/SysML-v2-Pilot-Implementation}} for Jupyter\footnote{\url{https://jupyter.org/}} with a lightweight compatibility layer. Since the original kernel is written in Java, it runs on the JVM in a separate process and communicates with .NET Interactive via JSON-RPC transparently from the user's perspective.
    \item Before moving to Journey, we built POLYGLOT using a custom middleware\footnote{\url{https://github.com/dotnet/interactive/blob/main/docs/kernels-overview.md\#middleware}} to intercept code submissions, run code analyzers, capture the produced events (such as the returned value), and provide custom magic commands\footnote{\url{https://github.com/dotnet/interactive/blob/main/docs/magic-commands.md}} in the context of mixed modelling and programming exercises \cite{POLYGLOT}.
    \item Following the philosophy of versatile delivery introduced in \hyperref[delivery]{chapter 1}, we have been experimenting with Amazon Alexa\footnote{\url{https://en.wikipedia.org/wiki/Amazon_Alexa}} as a frontend for learning activities such as lessons and quizzes.
\end{itemize}

\subsection{Journey}
\paragraph{}
Journey\footnote{\url{https://github.com/dotnet/interactive/tree/main/src/Microsoft.DotNet.Interactive.Journey}} is an open-source library, now part of the main .NET Interactive product, that enables the creation of open-ended interactive learning experiences. It is based on a few simple concepts:
\begin{itemize}
    \item A lesson is a logical equivalent to an exercise or a group of exercises. Setup and cleanup code can be defined, for example, to load libraries or data for the student transparently. A lesson is composed of challenges that are not limited in number or sequence.
    \item A \textbf{challenge} represents a step of an exercise. Challenges capture events produced by the kernels and memorize the submission history. Each challenge has its own customizable rules to perform validation tasks. Like lessons, challenges can have setup code, but they can also have post-submission code to perform arbitrary tasks (e.g. drawing a chart with the data filtered by the student) and to decide what challenge to execute next based on the submission's and rules' results.
    \item A \textbf{rule} represents a characteristic about the student's answer to be verified. It has a piece of code to perform validation tasks and decide if the student's submission satisfies the required property or not. Rules do not only represent mandatory characteristics; they can also represent additional desired properties that may differentiate a good answer from a great one. Each rule either fails with customized feedback or succeeds with a message.
\end{itemize}

\section{Interactive notebooks \& Amazon Alexa}
\subsection{Interactive notebooks}
\paragraph{}
"The idea of a notebook is to have an interactive document that freely mixes code, results, graphics, text and everything else"\footnote{\url{https://writings.stephenwolfram.com/2016/09/how-to-teach-computational-thinking/}} (Steven Wolfram). Most notebooks implementations are usually web-based applications and take advantage of web technologies to provide that flexibility and portability. Notebook documents are composed of cells that contain code snippets and produce output (or multiple outputs) and can execute independently of one another. These programming tools are already widespread in computational-based domains like data science but are gaining traction in education to support a wide variety of pedagogical scenarios\footnote{\url{https://www.epfl.ch/education/educational-initiatives/jupyter-notebooks-for-education/teaching-and-learning-with-jupyter-notebooks/}}.

\paragraph{}
.NET Interactive integrates seamlessly with notebook environments like Jupyter and VS Code. Those notebook environments send commands to the .NET Interactive composite kernel\footref{foot:dotnetKernelOverview} and react to the produced events and output. However, VS Code also exposes a notebook kernel that can receive commands and produce events. This addition enables a deeply integrated interaction with the editor. Journey uses this peculiarity to create cells dynamically as the student progresses through challenges. Thus, the resulting experience is not defined a priori but built incrementally and revealed to the student only when needed so that learners are not flooded with information at the beginning and can better focus on relevant content.

\subsection{Alexa}
\label{sec:alexa}
\paragraph{}
Alexa\footnote{\url{https://developer.amazon.com/en-US/alexa}} is a virtual assistant made by Amazon, capable of carrying out various automated tasks like controlling smart home devices or providing real-time information on sports or news. Users can interact with Alexa via a voice-based conversational user interface that third-party services can extend through apps called "skills".

\begin{figure}[htbp]
    \centering
    \includegraphics[width=\textwidth]{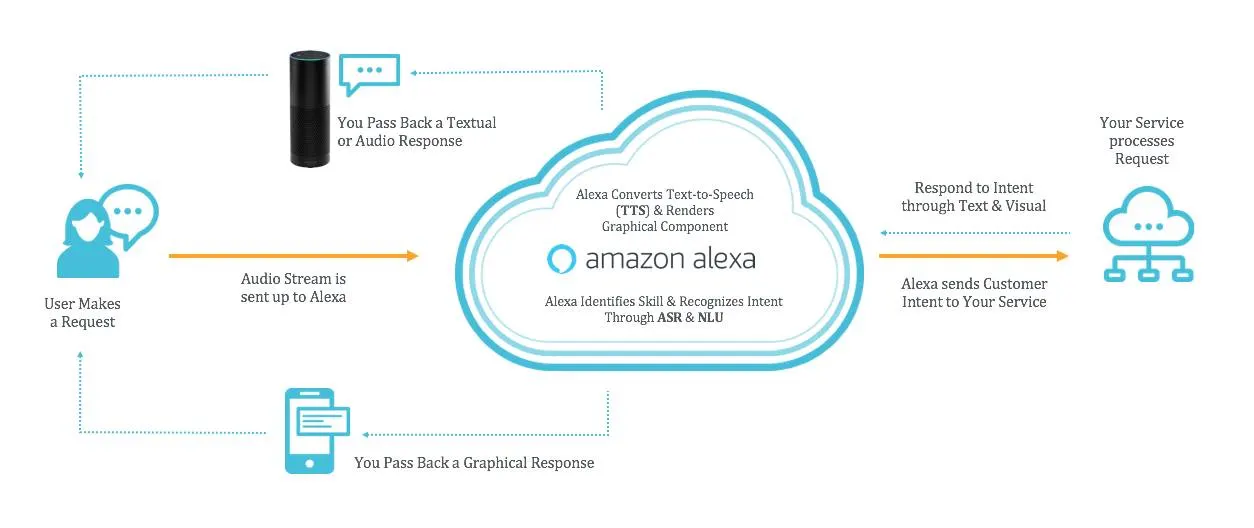}
    \caption[]{Alexa request flow\footnotemark}
    \label{fig:alexaFlow}
\end{figure}
\footnotetext{Image from \url{https://trailhead.salesforce.com/en/content/learn/modules/alexa-development-basics/get-started-with-alexa}}

\paragraph{}
When a user invokes Alexa by saying the wake word (i.e. "Alexa") followed by a sentence, the device sends the captured audio, called utterance, to the cloud. Developers provide a number of sample utterances to help Alexa map invocation to intents. There, speech recognition models and natural language understanding techniques are used to route the request to the correct intent handler registered by a skill. Intent handlers then process the created intent and produce a response that may be textual or/and video (for those devices that support it). Last, a text-to-speech engine transforms the response and speaks it to the user. 

%% file: chapters/chapter_4/chapter_4.tex
\chapter{Implementation}
\label{chapter:four}
\section{Architecture overview}
\paragraph{}
As stated previously, the project aims to provide an open, content-agnostic and extensible framework for designing and consuming adaptive and gamified learning experiences. We imagine a student experience entirely tailored to their needs and choices. For example, we think students should be able to do some lessons and quizzes with Alexa, switch to VS Code to do some coding activities, and then move to another frontend to do something else, all without friction. That is why we exploit the flexibility of \texttt{.NET Interactive} in POLYGLOT's execution engine. Students' interactions with external tools occur through \texttt{Adapters} that bind actions on the student frontend to \texttt{.NET Interactive} commands and bind \texttt{.NET Interactive} events to a supported output (e.g. audio for Alexa) with custom formatters.

\begin{figure}[htbp]
    \centering
    \includegraphics[width=\textwidth]{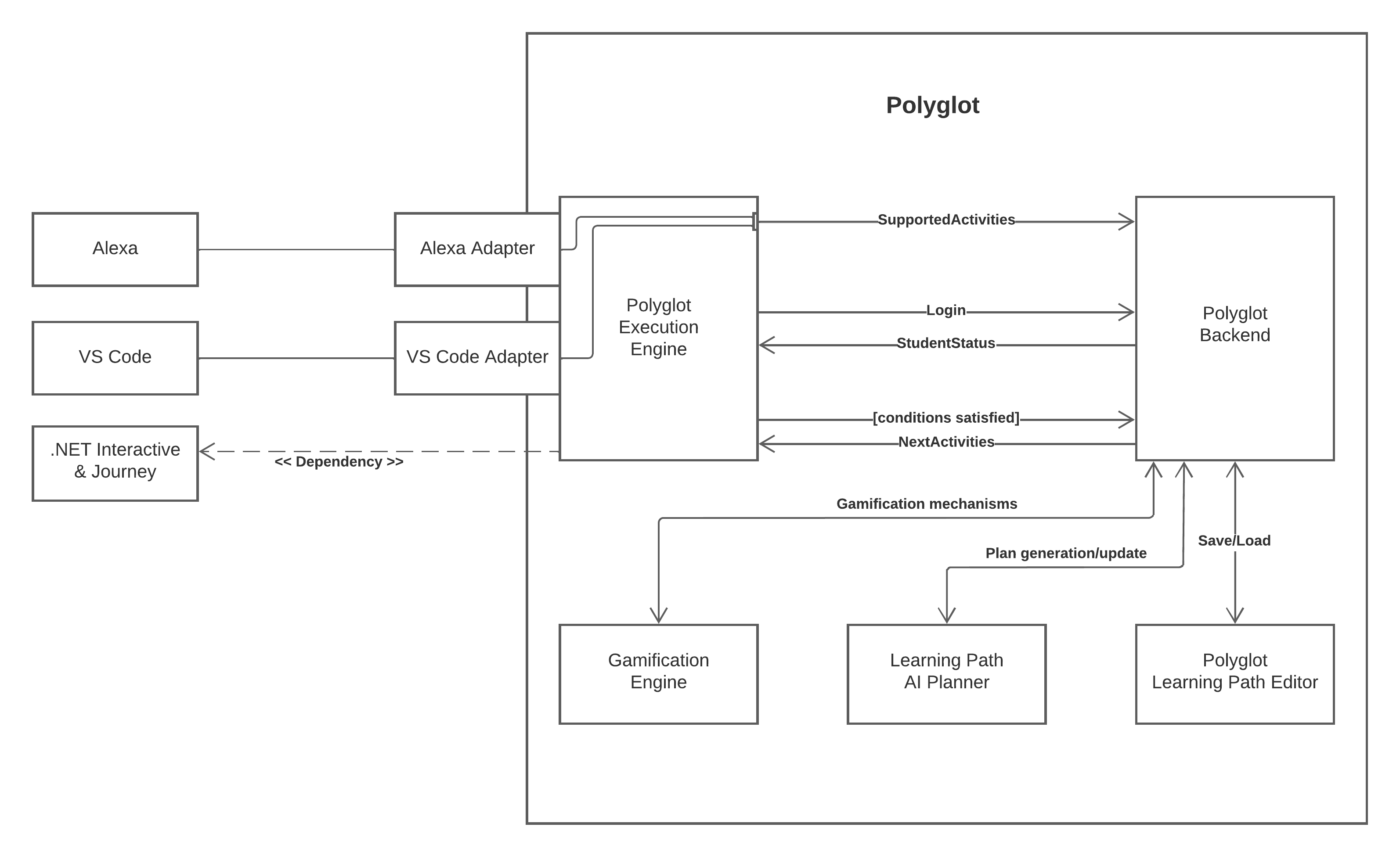}
    \caption[]{POLYGLOT architectural overview}
    \label{fig:polyglotArchitecture}
\end{figure}

\paragraph{}
The word "adaptive" in adaptive learning paths means being able to adjust the students' needs, like assigning exercises based on their previous responses or the capabilities of the platform they are on. This adaptation is allowed by the collaboration between the \texttt{Execution Engine} and the \texttt{Backend}. The former handles the students' submissions and validates their responses, while the latter uses the results of the validation phase to assign the next activity in the learning path.
\paragraph{}
The two most important aspects of an educational platform are \textit{simplicity in the content creation} process and \textit{content availability} to students for consumption and teachers for reuse. Designing and realizing learning fragments for such a variegated learning experience might seem a daunting task at first. That is why POLYGLOT includes a \texttt{Learning Path Editor} heavily focused on the teachers' experience. Its visual-editing capabilities and integration with the rest of the platform allow the creation of ready-to-use learning fragments with ease, thanks to the abstractions provided.
\paragraph{}
The \texttt{Gamification Engine} and the \texttt{AI Planner} are already working with internal prototypes. However, they have not yet been integrated even though the architecture is ready.

\section{Teacher design tool}

\begin{figure}[htbp]
    \centering
    \includegraphics[width=\textwidth]{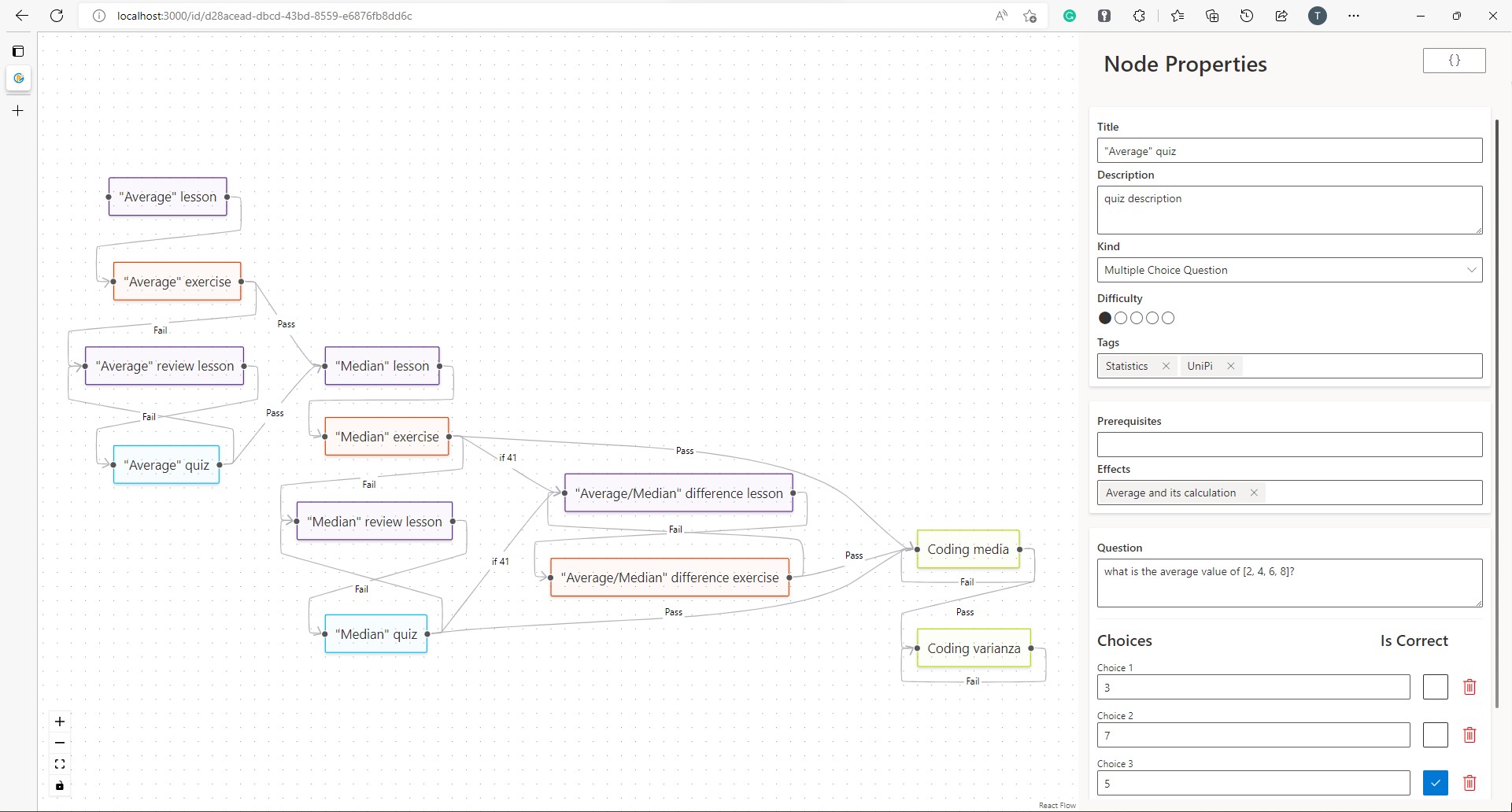}
    \caption[]{Teacher design tool for learning fragments}
    \label{fig:teacherFrontend}
\end{figure}

\paragraph{}
The teacher design tool, shown in \hyperref[fig:teacherFrontend]{figure 4.2}, is a visual editor for learning fragments (as described in \hyperref[sec:learningFragments]{section 2.2}). It aims to simplify the creation of said fragments by providing useful abstractions and powerful composition tools. The UI comprises two main elements: the \textit{drawing area} and the \textit{properties panel}. The drawing area is the core of the visual editing experience. Nodes can be added, connected, and rearranged with simple clicks or drag-and-drop interactions. Nodes and edges themselves can be interactive components that augment the visual editing experience. The properties panel, instead, provides fine-tuning tools to define the activities and the links between them.

\paragraph{}
To define the learning fragment, a teacher can create new activities with a right-click on the canvas, change their type and parameters from the properties panel, and connect them together by dragging from the handle on the source node to the handle on the destination node. By clicking on an edge, the teacher can edit the link condition by choosing from a list of existing abstractions or by writing their own validation code.

\begin{figure}[htbp]
    \centering
    \includegraphics[width=\textwidth]{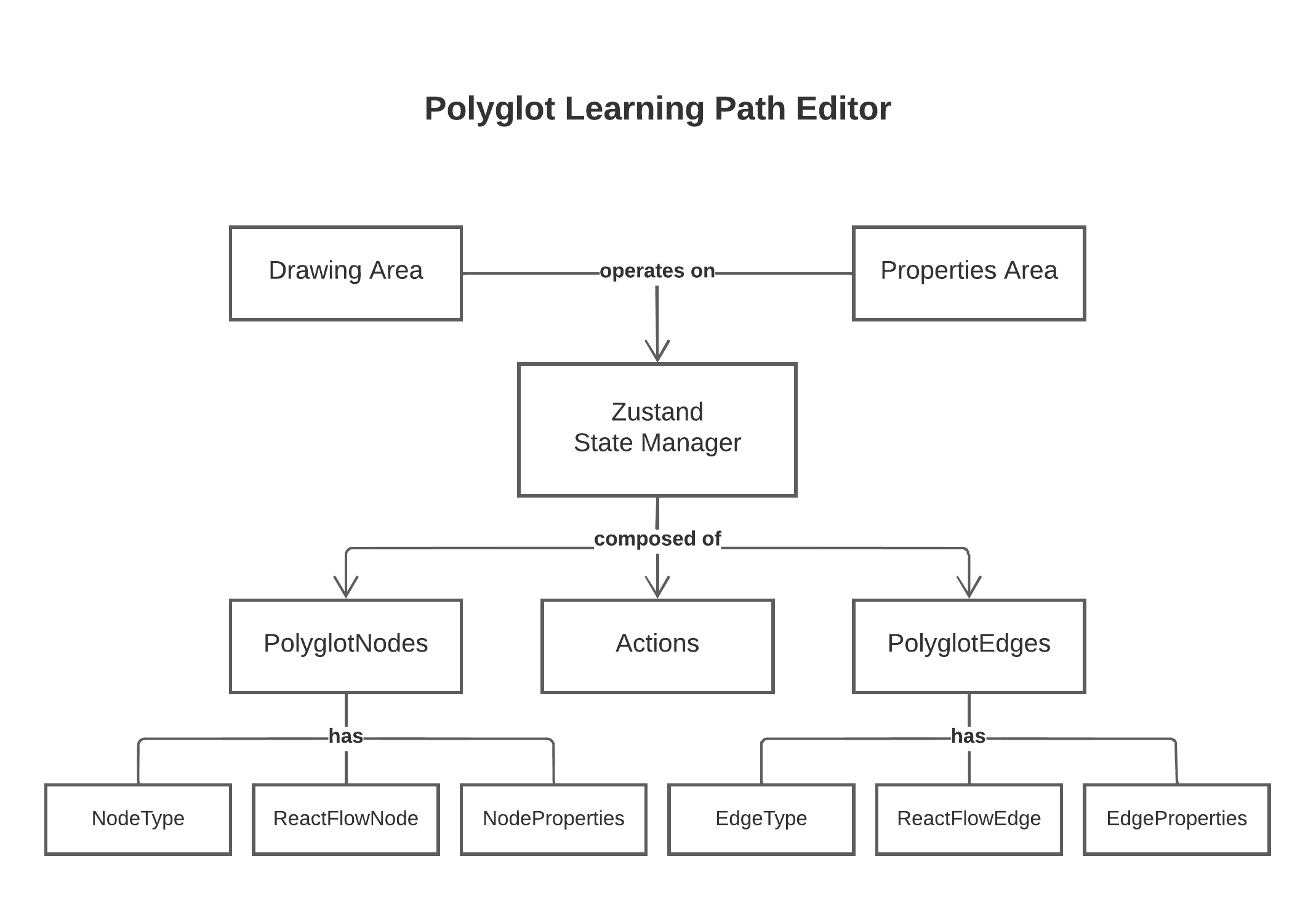}
    \caption[]{Teacher design tool architecture overview}
    \label{fig:teacherFrontendArchitecture}
\end{figure}

\paragraph{}
The tool is implemented using web technologies with React\footnote{\url{https://reactjs.org/}} and TypeScript\footnote{\url{https://www.typescriptlang.org/}} with a particular focus on flexibility and extensibility. Its design allows it to be part of a more comprehensive learning management tool. The drawing area and the properties panel are different abstract views that operate on the same data differently. Both components manipulate the same PolyglotElements (i.e. nodes and edges) by working on the shared state via Zustand\footnote{Zustand is a lightweight, unopinionated state-management library for React \url{https://github.com/pmndrs/zustand}} actions (see \hyperref[fig:teacherFrontendArchitecture]{figure 4.3}).

\subsection{The drawing area}
\begin{figure}[htbp]
    \centering
    \includegraphics[width=\textwidth]{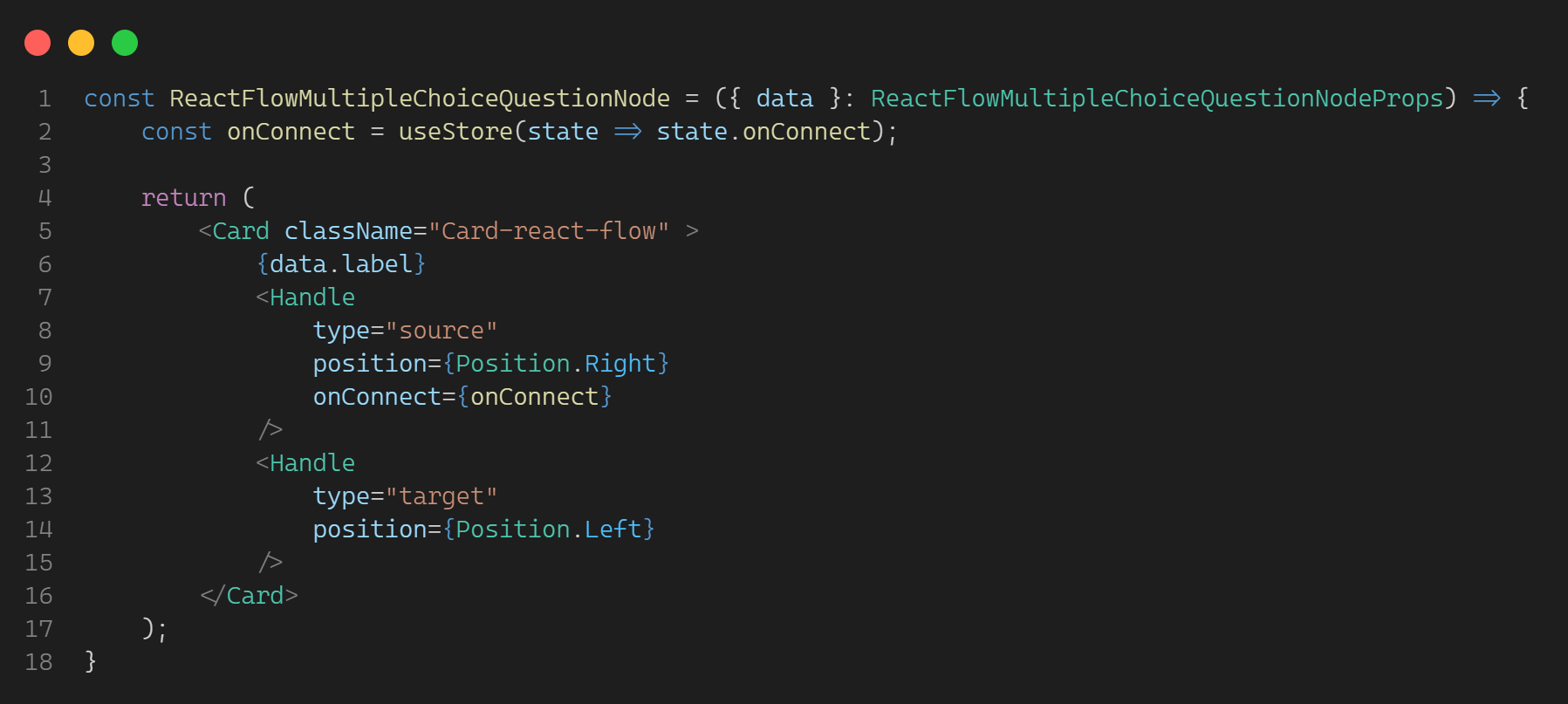}
    \caption[]{ReactFlowNode component for the multiple choice question node}
    \label{fig:reactFlowMultipleChoiceNode}
\end{figure}

\paragraph{}
The drawing area builds its functionalities upon ReactFlow\footnote{\url{https://reactflow.dev/}}, a React library for building node-based editors and interactive diagrams. As mentioned previously, nodes are interactive components themselves, personalizable in style and functionalities through custom ReactFlowNode component implementations (the node UI component in ReactFlow), as already happens for existing nodes (see \hyperref[fig:reactFlowMultipleChoiceNode]{figure 4.4} for a sample implementation). Handles\footnote{\url{https://reactflow.dev/docs/api/nodes/handle/}} are the component responsible for validating and accepting connections between nodes (more about validation in the \hyperref[sec:futureWorks]{future works section}). Connecting two nodes creates a new ReactFlowEdge: a customizable interactive component not significantly different from the ReactFlowNode.

\paragraph{}
ReactFlowNodes and ReactFlowEdges are one of the three components that define a PolyglotNode or PolyglotEdge. They may display some properties relative to their respective node or edge, so they might depend on a portion of the shared state and should be updated and re-rendered when the dependency changes. However, those components are not just consumers: even simple actions like moving a node or creating a connection modify the state, and those changes should be reflected everywhere they are needed.

\subsection{The properties panel}
\paragraph{}
The properties panel is the primary configuration tool for all PolyglotElements, including the fragment itself. Regardless of their type, all nodes share a set of fundamental characteristics and the same holds for edges. The properties component, therefore, contains two subcomponents: one for nodes' general properties and the other for type-specific ones. The same subdivision happens for edges. When the user selection changes, the sidebar is automatically updated and filled with both subcomponents according to the selected element (i.e. node or edge) and its type.

\begin{figure}[htbp]
    \centering
    \includegraphics[width=\textwidth]{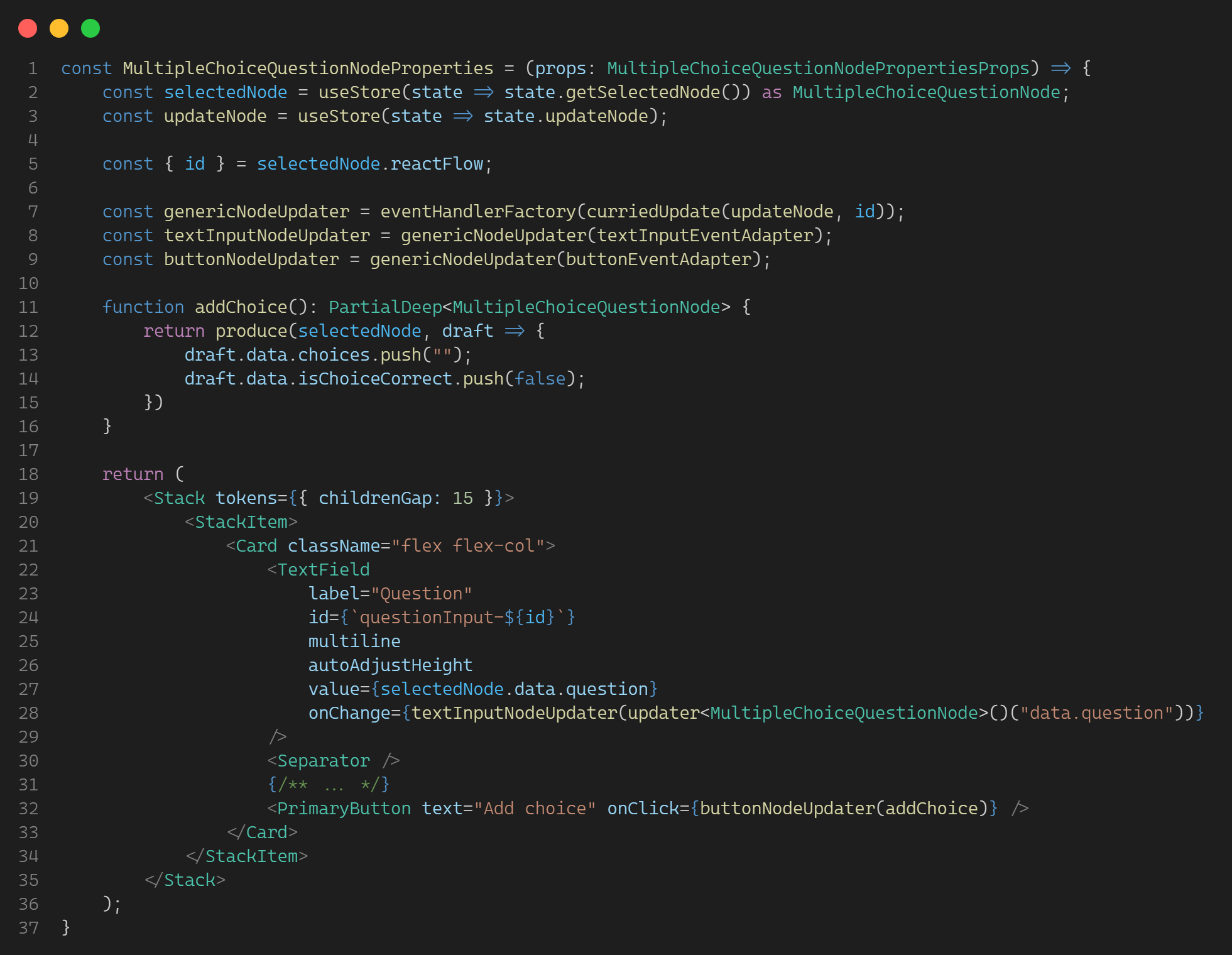}
    \caption[]{Properties component for the multiple choice question node}
    \label{fig:propertiesMultipleChoiceNode}
\end{figure}

\paragraph{}
Maintaining a consistent state is necessary to ensure the correctness of the existing elements and their relationships. On the contrary, allowing arbitrary operations on the state is also necessary to provide flexibility and extensibility. Zustand adopts an immutable state model\footnote{\url{https://en.wikipedia.org/wiki/Immutable_object}} where actions are not allowed to modify the state directly, but a copy must be created first. This approach has significant advantages but adds boilerplate to those actions that modify the state, thus increasing the error margin if done by hand. However, updates are made trivial by combining customized composable type-safe input handlers and the use of Immer\footnote{\url{https://immerjs.github.io/immer/}} (a package to work with immutable objects conveniently) to create strongly typed PartialDeep\footnote{A Partial type is a type whose properties are all made optional. A PartialDeep type extends the concept of Partial by recursively making optional the properties inside nested objects. For further information check: \url{https://github.com/sindresorhus/type-fest/blob/79f6b6239b270abc1c1cd20812a00baeb7f9fb57/source/partial-deep.d.ts\#L36}} updates that will be deeply merged into the existing state with the exposed actions. The result of this approach is present in \hyperref[fig:propertiesMultipleChoiceNode]{figure 4.5} on lines 7-16, 28 and 32.

\subsection{PolyglotElement processing}
\label{mentionValidationCode}
\begin{figure}[htbp]
    \centering
    \includegraphics[width=\textwidth]{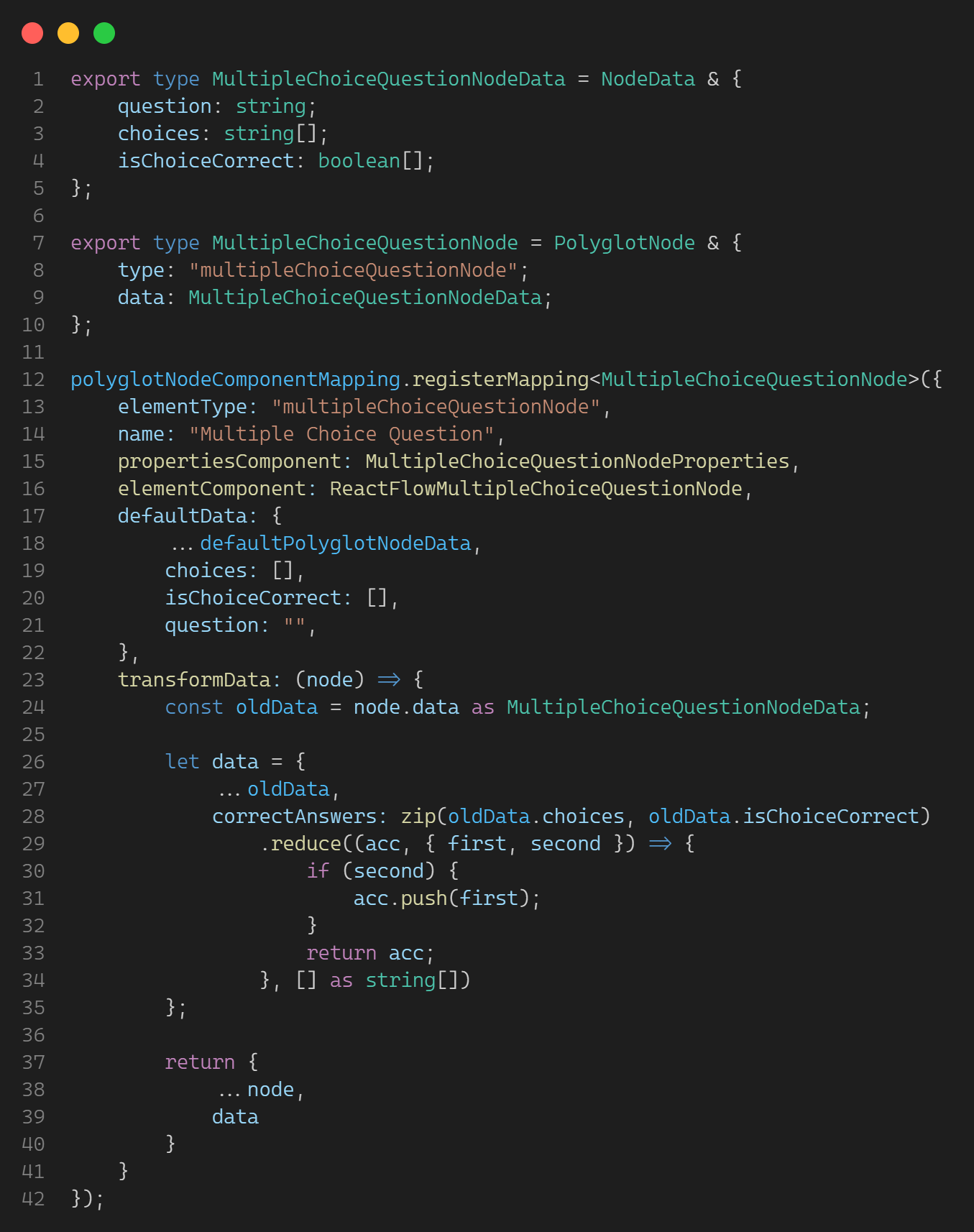}
    \caption[]{Multiple choice question node type}
    \label{fig:typeMultipleChoiceNode}
\end{figure}

\paragraph{}
The teacher design tool tries to achieve two opposite goals: providing useful abstraction to allow teachers to define the fragments naturally while remaining concrete enough to allow students to consume them without further intervention. This dual goal is especially challenging to achieve with edges where existing types allow teachers to define links abstractly with clauses such as "pass" or "fail", but, as explained in \hyperref[motivatingScenarioSection]{section 2.2.1}, not all activities are suitable for that abstract condition. Activities and connections must then undergo a programmable transformation phase to be converted into their executable counterparts. Implementation details can be generated behind the scenes to bridge the gap between the abstract world of activities and connections and its concrete executable form. Furthermore, some needed values may depend on the combination of others, so generating them guarantees consistency. For example, edges correspond to \texttt{Journey} rules in the execution phase. This correspondence is achieved by generating the validation code in this transformation phase. \hyperref[fig:typeMultipleChoiceNode]{Figure 4.6} shows an example transformation to generate additional node attributes.

\subsection{Adding node and edge types}
\paragraph{}
The whole architecture and the individual components have been carefully designed to minimize the knowledge and expertise needed to extend the platform. Developers can easily create additional custom nodes and edges just by adding two (or optionally three) components:
\begin{itemize}
    \item A node (or edge) type that extends the base PolyglotNode type or every other preexisting subtype (\hyperref[fig:typeMultipleChoiceNode]{figure 4.6}).
    \item A NodeProperties (or EdgeProperties) component for the properties panel to edit type-specific data (\hyperref[fig:propertiesMultipleChoiceNode]{figure 4.5}).
    \item An optional ReactFlowNode (or ReactFlowEdge) to customize the appearance and functionalities in the drawing area (\hyperref[fig:reactFlowMultipleChoiceNode]{figure 4.4}).
\end{itemize}
Those are virtually the only components needed in the entire POLYGLOT architecture to add custom activities. \texttt{.NET Interactive} formatters can further customize the appearance of such activities in students' frontends. The source code is available on GitHub at \url{https://github.com/polyglot-edu/node-editor}.

\section{Execution engine}
\begin{figure}[htbp]
    \centering
    \includegraphics[width=\textwidth]{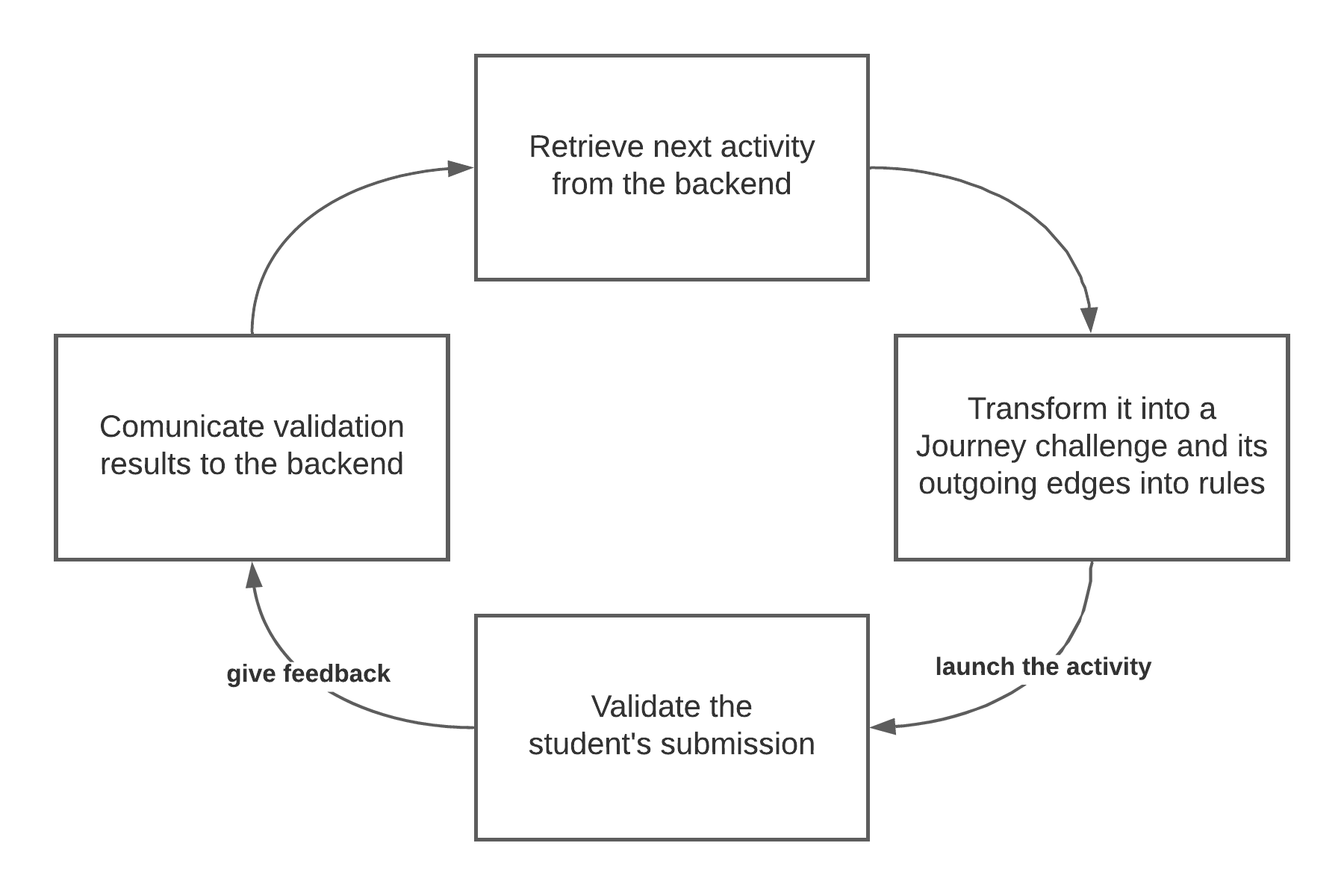}
    \caption[]{POLYGLOT execution engine lifecycle}
    \label{fig:engineLifecycle}
\end{figure}

\paragraph{}
This component constitutes the core of the student experience. It is responsible for handling user interactions and validating students' submissions. Its straightforward lifecycle (\hyperref[fig:engineLifecycle]{figure 4.7}) is designed to exploit \texttt{Journey}'s strengths by using it in a semi-controlled fashion. When the assigned activity and its outgoing edges are retrieved from the backend, they undergo a transformation phase where the activity is converted into a \texttt{Journey} challenge and edges into \texttt{Journey} rules. This process is essential because it allows running the validation code mentioned in \hyperref[mentionValidationCode]{section 4.2.3}. Once converted, the challenge is ready to be launched. First, the setup code gets executed, and then the activity is "shown" to the student in their frontend, thanks to \texttt{.NET Interactive} formatters. When the active frontend adapter produces an event, \texttt{Journey} intercepts it and runs the registered rules against the student submission. POLYGLOT then sends the result of this validation phase back to the backend that will use this additional knowledge to suggest the next suitable activity.

\subsection{The conversion process}
\paragraph{}
Abstractly, a connection between two activities means that if a property on the source activity is satisfied, the student is allowed to pass to the destination activity. This process needs three components to work correctly:
\begin{itemize}
    \item information on the source activity
    \item information on the student submission
    \item information on what property to enforce and how to do it
\end{itemize}

\paragraph{}
Suppose a quiz activity has an outgoing edge that enforces the property "pass". In this context, "pass" means that the student's answer must be among the correct solutions to the exercise. If we were to translate it into code, we would write a function that takes three inputs (the components mentioned above) and returns a boolean representing the validation result.

\begin{figure}[htbp]
    \centering
    \includegraphics[width=\textwidth]{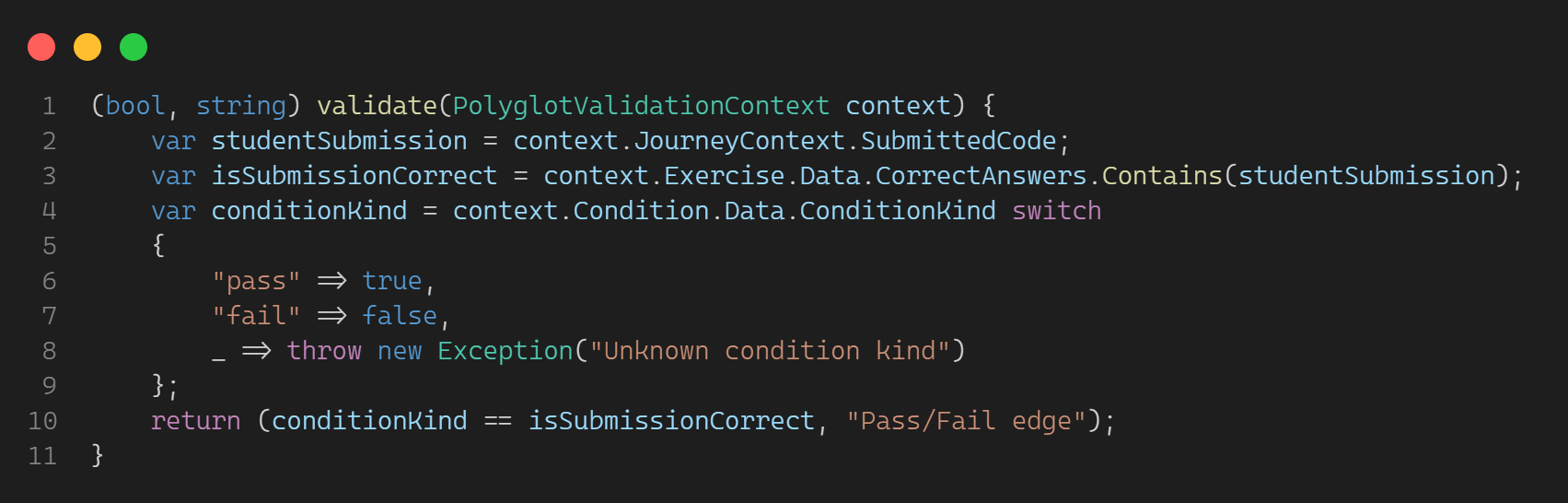}
    \caption[]{Pass/Fail abstract condition implementation}
    \label{fig:passfailCode}
\end{figure}

\paragraph{}
The actual code (\hyperref[fig:passfailCode]{figure 4.8}) is not too far from the previous description. It takes a PolyglotValidationContext as input and returns a pair (result, feedback). The PolyglotValidationContext type contains the three essential components:
\begin{itemize}
    \item a \texttt{Journey} RuleContext that holds information on the student submission
    \item an Exercise that contains information on the source activity
    \item and a Condition that contains information on the edge (e.g. if the condition is pass or fail)
\end{itemize}

\paragraph{}
\CSharp\ is a compiled language, so the condition code must be compiled at runtime with the CSharpScript API\footnote{The scripting APIs enable .NET applications to instatiate a \CSharp\ engine and execute code snippets against host-supplied objects. \url{https://github.com/dotnet/roslyn/blob/daab8026f666f07fb43aef9126a98e9968a355b6/docs/wiki/Scripting-API-Samples.md}} before \texttt{Journey} can use it in rules. \CSharp\ is also a statically typed language; thus, the deserialization of Exercise and Condition requires their type to be known in advance. Given our extensibility goal, we overcome this limitation using DLR's\footnote{DLR is .NET's Dynamic Language Runtime \url{https://docs.microsoft.com/en-us/dotnet/framework/reflection-and-codedom/dynamic-language-runtime-overview}} ExpandoObject\footnote{An ExpandoObject represents an object whose members can be dymanically modified at runtime \url{https://docs.microsoft.com/en-us/dotnet/api/system.dynamic.expandoobject?view=net-6.0} \url{https://www.newtonsoft.com/json/help/html/T_Newtonsoft_Json_Converters_ExpandoObjectConverter.htm}} to deserialize the exercise and the condition data into a dynamically typed variable\footnote{\url{https://docs.microsoft.com/en-us/dotnet/csharp/programming-guide/types/using-type-dynamic}}. This method allows us to run arbitrary activities (even newly created ones) without needing the execution engine to know their existence.

\paragraph{}
The execution engine is available as a NuGet package along with the SysML kernel and various \CSharp\ and SysML analyzers. The source code is available on GitHub at \url{https://github.com/polyglot-edu/runtime}.

\pagebreak
\section{Other components}
\subsection{Backend}
\paragraph{}
At the moment, the backend is not a significantly complex component. Its main task is to support the teacher's design tool and the execution engine by providing RESTful APIs to operate on the learning fragments.
It is implemented with TypeScript and Express\footnote{\url{https://expressjs.com/}}, and the source code is available on GitHub at \url{https://github.com/polyglot-edu/backend}.

\subsection{Alexa adapter}
\paragraph{}
The Alexa integration with POLYGLOT is composed of two major components: the Alexa skill and an ASP.NET server hosted on Azure. As explained in \hyperref[sec:alexa]{section 3.3.2}, Alexa transforms utterances into custom-defined intents. Our intent handler sends the content of the intent slot\footnote{\url{https://developer.amazon.com/en-US/docs/alexa/custom-skills/create-the-interaction-model-for-your-skill.html\#intents-and-slots}} to the ASP.NET server. Once there, the user request is converted into a SubmitCode command, sent to the appropriate kernel, and executed as expected by \texttt{.NET Interactive}, \texttt{Journey} and POLYGLOT's execution engine. The events produced are sent back to Alexa, which reads their content to the user.

\begin{figure}[htbp]
    \centering
    \includegraphics[width=\textwidth]{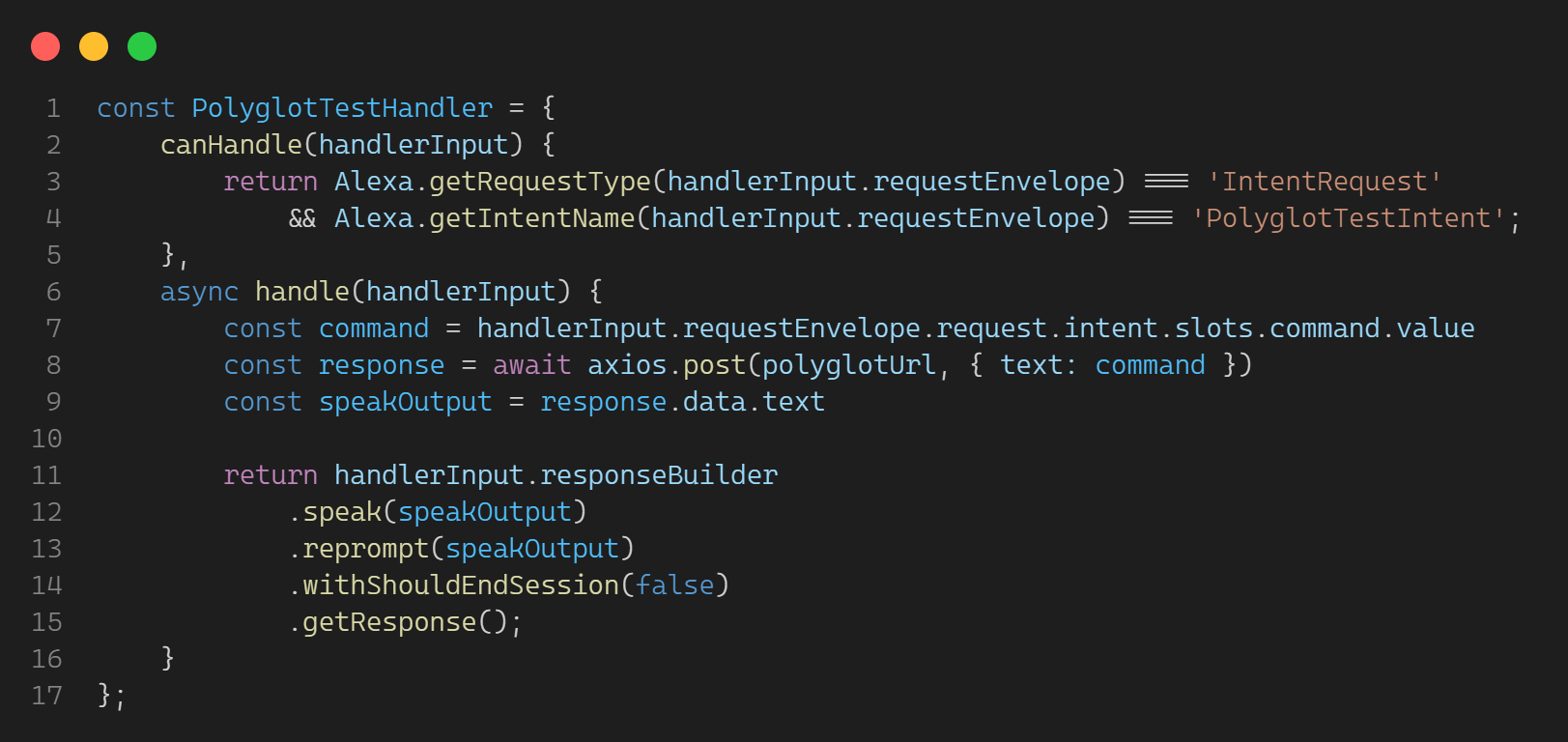}
    \caption[]{Simplified implementation of Alexa's intent handler}
    \label{fig:alexaCode}
\end{figure}

%% file: chapters/chapter_5/chapter_5.tex
\chapter[Conclusions and Future Directions]{Conclusions and\\Future Directions}
\label{chapter:five}
\label{sec:futureWorks}
\paragraph{}
Adaptive education still remains an open research topic at the intersection of several different domains like cognitive sciences and informatics. We have identified some directions to explore in future collaborations:

\begin{itemize}
    \item \textbf{Abstract activities} are at the core of our formalization. Extending POLYGLOT to include them is our highest priority. To do that, we need to include them in the editor, design a tool for defining goals, and integrate the AI planner to include the runtime refinement and harness the power of abstract activities. We are already working on all these tasks with the help of another intern at FBK.
    \item Runtime refinement is one of the most powerful concepts in the entire framework. However, it is most effective only if there is a substantial amount of fragments to choose from. We think that creating an \textbf{open fragment database} would benefit students (with better refinement and high-quality educational content) and teachers who can focus on the big picture and let the platform adjust the details.
    \item To further extend the previous point, \textbf{leveraging open educational resources with NLP techniques} for automatic content annotation would significantly contribute to the database. This task is one of the main goals of Project Encore\footnote{\url{https://grial.usal.es/encore}}, an ERASMUS\texttt{+} project that involves multiple universities and companies.
    \item Allowing students to use \textbf{arbitrary frontends} for their tasks is indeed challenging. This support is still experimental and needs more work to ensure the quality of the learning experience. Furthermore, not all frontends are suitable for all kinds of tasks: for example, using Alexa to do a coding exercise might not be the best idea. We need to further define what "suitable" means and implement something similar to a content negotiation mechanism.
    \item To easily define the fragments, POLYGLOT provides some basic abstractions for connections such as "Pass/Fail". As explained previously, this abstraction may be correct for some exercises but not applicable to others. We want to further explore this topic to enhance the correctness of existing fragments, \textbf{validate connections} to prevent the creation of erroneous links, and define other composable abstractions.
    \item We plan to add \textbf{support for existing learning management systems}\footnote{\url{https://en.wikipedia.org/wiki/Learning_management_system}} (e.g. Moodle LMS) in the same modular fashion as we are doing with students' frontends. This extension would allow incremental integration of parts of POLYGLOT within existing learning infrastructures and increase the adoption of adaptive learning technologies.
    \item The combined evolution of the engagement and learning systems described in \hyperref[sec:gamification]{section 2.3} is still an open research problem. Gamification mechanisms are effective only when they fit the learning path perfectly. Usually, it is the gamification designer's duty to design the game narrative and the other game elements to achieve that fit. However, the power of adaptive learning paths comes from the runtime refinement; therefore, paths are not available upfront. The engagement system must then evolve through an \textbf{automatic calibration} phase to the mechanisms to the underlying activities.
    \item The learning path planning is crucial for effective refinement. We aim to enhance the planner by leveraging \textbf{machine learning techniques using learners' behavioural data}. Our goal is to improve contextual activity suggestions (and hopefully feedback), considering non-standard parameters like the capabilities of the interface the student is using. 
\end{itemize}

\paragraph{}